\documentclass[final,3p,times]{elsarticle}
\usepackage{graphicx}
\usepackage{amssymb}
\usepackage{amsmath}
\usepackage{subfig}
\usepackage{graphicx}
\usepackage{url}
\usepackage{algorithm}
\usepackage{algpseudocode}
\usepackage{multirow}
\usepackage{wasysym}
\usepackage{marvosym}
\usepackage{color, xcolor, colortbl}
\usepackage{array, booktabs}
\usepackage{xspace}
\usepackage{enumitem}
\usepackage{bm}
\definecolor{light}{rgb}{0.8,0.8,0.8}
\definecolor{medium}{rgb}{0.6,0.6,0.6}
\definecolor{dark}{rgb}{0.4,0.4,0.4}
\definecolor{darkmed}{rgb}{0.3,0.3,0.3}
\definecolor{darkest}{rgb}{0.2,0.2,0.2}
\definecolor{Black}{rgb}{0,0,0}
\definecolor{White}{rgb}{1,1,1}
\definecolor{lightpurple}{rgb}{0.78823,0.709803,0.74509}
\definecolor{lightpurpletext}{rgb}{0.788235,0.5529411,0.658823}
\definecolor{skyblue}{rgb}{0.80392,0.866666,0.92941}
\definecolor{skybluetext}{rgb}{0.61568627,0.7647058,0.913725}
\definecolor{darkgreen}{rgb}{0.3137254,0.458823,0.18431}
\definecolor{foliagegreen}{rgb}{0.188,0.415,0.105}
\definecolor{steelbluegrey}{rgb}{0.1961,0.2353,0.2392}
\definecolor{highlightblue}{rgb}{0.4078,0.6431,0.85}
\definecolor{matlabblue}{rgb}{0,0.2705,0.85}
\definecolor{darkred}{rgb}{0.8,0.1725,0}
\definecolor{fireenginered}{rgb}{0.505,0.1411,0}
\definecolor{darkpurple}{rgb}{0.6431,0.3137,0.8509}
\definecolor{gaylordpurple}{rgb}{0.416,0.204,0.549}
\definecolor{deludedorange}{rgb}{0.7409,0.4392,0}
\definecolor{darksalmon}{rgb}{0.9137,0.411,0.706}
\newcommand{\secref}[1]{Section \ref{#1}}
\newcommand{\fref}[1]{Fig.~\ref{#1}}

\renewcommand{\eqref}[1]{Eq.~(\ref{#1})}
\newcommand{\eref}[1]{(\ref{#1})}

\newcommand{\mat}[1]{\textrm{\textbf{#1}}}

\newcommand{\psif}{\bm{\psi} (\bm{r}, \bm{\Omega})}
\newcommand{\psifd}{\bm{\psi} (\bm{r}, \bm{\Omega}')}

\newcommand{\xten}[1]{$\times$ 10$^{\text{#1}}$}

\newcolumntype{a}{>{\columncolor{light}}c}
\newcolumntype{b}{>{\columncolor{skyblue}}c}

\begin{document}

\begin{frontmatter}



%~~~~~~~~~~~~~~~~~~~~~~~~~~~~~~~~~~~~
\title{Angular adaptivity with spherical harmonics for Boltzmann transport}
\author[AMCG]{S. Dargaville}
\ead{dargaville.steven@gmail.com}
\author[QMU]{A.G. Buchan}
\author[AWE,AMCG]{R.P. Smedley-Stevenson}
\author[AMEC,AMCG]{P.N. Smith}
\author[AMCG]{C.C. Pain}
\address[AMCG]{Applied Modelling and Computation Group, Imperial College London, SW7 2AZ, UK}
\address[QMU]{School of Engineering and Material Sciences, Queen Mary University of London, E14 NS, UK}
\address[AWE]{AWE, Aldermaston, Reading, RG7 4PR, UK}
\address[AMEC]{ANSWERS Software Service, Wood PLC, Kimmeridge House, Dorset Green Technology Park, Dorchester, DT2 8ZB, UK}
%~~~~~~~~~~~~~~~~~~~~~~~~~~~~~~~~~~~~
\begin{abstract}
This paper describes an angular adaptivity algorithm for Boltzmann transport applications which uses P$_n$ and filtered P$_n$ expansions, allowing for different expansion orders across space/energy. Our spatial discretisation is specifically designed to use less memory than competing DG schemes and also gives us direct access to the amount of stabilisation applied at each node. For filtered P$_n$ expansions, we then use our adaptive process in combination with this net amount of stabilisation to compute a spatially dependent filter strength that does not depend on \textit{a priori} spatial information. This applies heavy filtering only where discontinuities are present, allowing the filtered Pn expansion to retain high-order convergence where possible. Regular and goal-based error metrics are shown and both the adapted P$_n$ and adapted filtered P$_n$ methods show significant reductions in DOFs and runtime. The adapted filtered P$_n$ with our spatially dependent filter shows close to fixed iteration counts and up to high-order is even competitive with P$^0$ discretisations in problems with heavy advection. 
\end{abstract}
%~~~~~~~~~~~~~~~~~~~~~~~~~~~~~~~~~~~~
%~~~~~~~~~~~~~~~~~~~~~~~~~~~~~~~~~~~~
%~~~~~~~~~~~~~~~~~~~~~~~~~~~~~~~~~~~~

\end{frontmatter}
%~~~~~~~~~~~~~~~~~~~~~~~~~~~~~~~~~~~~
%~~~~~~~~~~~~~~~~~~~~~~~~~~~~~~~~~~~~
%~~~~~~~~~~~~~~~~~~~~~~~~~~~~~~~~~~~~
%-----------------------------
\section{Introduction}
\label{sec:Introduction}
The Boltzmann transport equation (BTE) is used to model the transport of neutral particles through an interacting medium and can be difficult to solve because of its mixed hyperbolic/parabolic nature and its (up to) seven-dimensional phase space. The large size of this phase space comes from the three spatial dimensions, two angular, one energy/frequency and one time dimensions. As such, many authors who use deterministic techniques to solve the BTE area turning to adaptive techniques to ensure accuracy in their discretisation while reducing the size of the resulting linear system. 

Performing adaptivity in the angular domain of the BTE is becoming more popular, as high angular resolution is required in many problems to resolve the streaming terms (i.e., when a particle propogates a long distance without interaction). Previously, in the AMCG we have investigated using several different angular discretisations with adaptivity across a range of different Boltzmann transport applications \cite{Buchan2008244, Goffin2014, Goffin2015, Goffin2015a, Adam2016a, Adam2016, Soucasse2017, Dargaville2019}. Recently \cite{Dargaville2019} we showed scalable angular adaptivity implemented matrix-free with Haar wavelets that demonstrated $\mathcal{O}(n)$ scaling in both runtime and memory usage on some test problems. This Haar wavelet discretisation in angle is equivalent to a P$^0$ FEM discretisation on the sphere, and like all non-rotationally invariant angular discretisations produces ray-effects in the scalar flux. As such, angular adaptivity becomes key, as resolution must be applied anisotropically on the sphere to smooth out ray-effects and to capture different streaming paths across space/energy.

The only rotationally invariant angular discretisation widely used in Boltzmann transport applications is a spherical harmonics discretisation of the sphere, otherwise known as P$_n$. This discretisation does not suffer from ray-effects, which would make it ideal for applications with heavy streaming, but its global span on the sphere and high-order nature give increasingly ill-conditioned systems. This is due to Gibbs oscillations when trying to capture the pathology of a hat function in angle due to heavy advection (or other discontinuities in angle, caused by material properties, etc).  As such, P$_n$ discretisations in angle are traditionally limited to problems with sufficient smoothness. 

Over the last several years however, filtered P$_n$ (we refer to these as FP$_n$) approximations have been introduced \cite{McClarren2010, Radice2013, Frank2016, Laboure2016b, Laboure2016, Laiu2018}, which aim to improve conditioning with discontinuities present. This follows the reasoning behind high-order spectral element methods in fields like aeronautics, where the high-order basis functions are filtered near discontinuities while still retaining high-order convergence in smooth regions. This however creates additional difficulties, as we must now decide what type of filter to apply, how much to filter and where a filter should be applied. If we filter too much in smooth areas for example, we destroy the convergence of the method. This leads naturally to making the filter strength spatially dependent. \cite{Laboure2016b, Laboure2016} are the only authors to do this, and showed an example in a 2D duct where they \textit{a priori} determined a region to filter heavily to improve their solutions.

This paper aims to combine both filtered P$_n$ discretisations and angular adaptivity in order to produce ray-effect free solutions to the BTE, with good conditioning in problems with heavy streaming. To begin, we produce a novel matrix-free angular adaptivity scheme for P$_n$ and FP$_n$ that uses both regular and goal-based error metrics. We then turn to the development of a spatially dependent filter strength for our FP$_n$ method. Previously we introduced a sub-grid scale FEM discretisation \cite{hughes_variational_1998, hughes_multiscale_2006, candy_subgrid_2008, buchan_inner-element_2010, Dargaville2019} that uses less memory than equivalent DG schemes. This scheme explicitly separates our solution into ``fine'' and ``coarse'' scales, with the fine scale representing the amount of stabilisation that is applied. 

Rather than use \textit{a priori} information to determine where to filter heavily, we use the net amount of stabilisation applied in each of our adapt steps to scale our filter strength across space. This automatically applies a heavy filter in areas with discontinuities, no matter the cause (e.g., material properties, heavy streaming, boundary conditions, or sources). This combination of spatially dependent filter and adapted FP$_n$ method is a powerful tool for achieving accurate, ray-effect free solutions in problems with varying degrees of smoothness. We show evidence of the success of our spherical harmonics adaptivity scheme on three example problems, with reductions in DOFs, runtime and memory consumption and highlight the benefit of combining this with filtering. 
% ~~~~~~~~~~~~~
\section{Boltzmann Transport Equation}
\label{sec:Boltzmann Transport Equation}
% ~~~~~~~~~~~~~
We begin with the  BTE and without loss of generality, we write the mono-energetic steady-state BTE in first-order form as
\begin{equation}
\bm{\Omega} \cdot \nabla_{\bm{r}} \psif + \Sigma_\textrm{t} \psif - S(\psif)  = S_{\textrm{e}}(\bm{r}, \bm{\Omega}),
\label{eq:bte}
\end{equation}
where $\psif$ is the angular flux in direction $\bm{\Omega}$, at spatial position $\bm{r}$. The macroscopic cross sections define the material that particles are moving through and $\Sigma_\textrm{t}$ is the total cross-section. The interaction/source terms have been separated into, $S$, which is dependent on $\psif$, and those which can be considered purely ``external'', $S_\textrm{e}$. In this work the only source term is the scattering from angle $\bm{\Omega}'$ into angle $\bm{\Omega}$ as particles interact with the medium they are propogating in, or
\begin{equation}
S(\psif) = \int_{\bm{\Omega}'} \Sigma_\textrm{s} (\bm{r}, \bm{\Omega}' \rightarrow \bm{\Omega}) \psifd \textrm{d}\bm{\Omega}',
\label{eq:scatter}
\end{equation}
where $\Sigma_\textrm{s}$ is the macroscopic scatter cross-sections. We now move to discussing spherical harmonics in more detail, along with an overview of the previous work on angular adaptivity.
% ~~~~~~~~~~~~~
\section{Background}
\label{sec:Background}
% ~~~~~~~~~~~~~
% ~~~~~~~~~~~~~
\subsection{Spherical harmonics}
\label{sec:pn}
% ~~~~~~~~~~~~~
We begin by writing a spherical harmonics expansion up to order $N$ of a function $f$, on the unit sphere, S$^1$, as 
\begin{equation}
f(\bm{\Omega}) = \sum_{l=0}^{N} \sum_{m=-l}^{l} f_{l,m} Y_{l,m}(\bm{\Omega}),
\label{eq:sph_harmonics}
\end{equation}
where $f_{l,m}$ are the expansion coefficients and $Y_{l,m}$ are the real, orthonormal spherical harmonics. If $f$ is smooth, \eref{eq:sph_harmonics} will converge to to $f$ with spectral accuracy as $N\rightarrow \infty$. If $f$ is not smooth, Gibbs oscillations will result and the approximation can result in negative solutions, which can be problematic in many applications. 
% ~~~~~~~~~~~~~
\subsection{Filtered spherical harmonics}
\label{sec:fpn}
% ~~~~~~~~~~~~~
One of the main motivations \cite{McClarren2010} considered in formulating a filtered P$_n$ method was to enforce positivity in a spherical harmonics expansion. We write the FP$_n$ method using the generalised notation of \cite{Radice2013}, where the P$_n$ expansion \eref{eq:sph_harmonics} with filtering becomes
\begin{equation}
f(\bm{\Omega}) = \sum_{l=0}^{N} \sum_{m=-l}^{l} \left[\sigma \left( \frac{l}{N+1} \right) \right]^s f_{l,m} Y_{l,m}(\bm{\Omega}),
\label{eq:fpn}
\end{equation}
where $\sigma(\eta)$ is a filter function that obeys certain properties \cite{Radice2013}, including having a value of 1 on the isotropic moment, and that $\sigma(\eta)$ only depends on $l$, making \eref{eq:fpn} rotationally invariant like \eref{eq:sph_harmonics}. A strength parameter, $s$ is also introduced into \eref{eq:fpn}. We note that as $N \rightarrow \infty$, or if $s=0$, \eref{eq:fpn} reduces to a normal spherical harmonics expansion. 

\cite{Radice2013} also showed that this formulation is equivalent to including a forward-peaked scattering operator to \eref{eq:bte}, making the implementation of a filtered P$_n$ method simple. This form simply adds a diagonal term to the discretised system, namely $-\Sigma_{\textrm{f}} \log(\sigma(l/m))$, where $\Sigma_{\textrm{f}}$ is a free parameter that is independent of the mesh spacing (and time step if applicable), can be made spatially dependent and as such can be considered as similar to a material property, which determines the ``filter strength'' ($\Sigma_{\textrm{f}}$ contains $s$). 

Hence we are left with two choices: what filter function, $\sigma(\eta)$, we chose and the value of $\Sigma_{\textrm{f}}$. We chose to use a Lanczos filter exclusively in this work given by $\sigma(\eta) = \sin(\eta)/\eta$, as \cite{Radice2013} shows this filter works well. A number of authors have investigated the effect of different filter functions \cite{Radice2013, Laboure2016, Laboure2016b, Frank2016}. \cite{Frank2016} in particular prove global convergence properties of filtered P$_n$ expansions, and determined as one may expect, that in smooth problems the order of the filter determines the rate of convergence. For non-smooth problems, it is the underlying smoothness of the transport problem that determines the rate of convergence. Given this work is largely focused on transport problems with a lack of smoothness, we do not examine the use of different filters. Finally we must determine the filter strength, $\Sigma_{\textrm{f}}$. To begin we assume $\Sigma_{\textrm{f}}$ is constant across the spatial mesh, but in \secref{sec:Spatially dependent filter} we make $\Sigma_{\textrm{f}}$ spatially dependent.
% ~~~~~~~~~~~~~
\subsection{Adaptivity with P$_n$}
\label{sec:aadapt}
% ~~~~~~~~~~~~~
Only a handful of authors have investigated using adaptivity with spherical harmonics in Boltzmann transport applications. We do not discuss other adapted angular discretisations, please see \cite{Goffin2015a, Dargaville2019} for a detailed overview.
\cite{Ackroyd1986, Ackroyd1988} first used different P$_n$ expansion orders across space in a 1D test problem, where the expansion order was determined \textit{a priori}.  \cite{Park2006, Park2009a} then used the even-parity form of the BTE (which is well suited to P$_n$ simulations as neither work in voids) and built a matrix-free combined space/angle adaptive algorithm that used goal-based metrics and showed reductions in total NDOFs applied. \cite{Rupp2011a} performed regular adaptivity with spherical harmonics in 1D, that showed a reduction in NDOFs. \cite{Goffin2014, Goffin2015a} used the first order form of the BTE and combined this with regular and goal-based error metrics to produce an adaptive P$_n$ algorithm, showing reductions in NDOFs across a range of 1D, 2D and 3D problems with energy dependence. The only example problem the authors showed with runtimes however, indicated their adaptive scheme did not improve upon the runtimes of an unadapted spherical harmonics discretisation. Finally \cite{Safarzadeh2015} formed a combined space/angle adaptivity algorithm using double spherical harmonics expansion and also showed a reduction in NDOFs.

This work is focused on applying angular adaptivity to the first order form of the BTE, which is much more difficult to solve than the even-parity form, given the resulting linear system is not SPD. Naturally the aforementioned authors have not investigated the use of P$_n$ expansions in voids given the conditioning problems of high order expansions, but unfortunately, even in the smooth problems presented, many of the authors do not show the runtimes of their adaptive methods. Our previous work \cite{Dargaville2019} focused heavily on the scalability of an angular adaptivity scheme based on Haar wavelets. This work follows on from \cite{Dargaville2019} and as such we discuss the scalability and runtime/memory consumption of our adaptive spherical harmonics scheme. 

The sparsity of the P$_n$ angular matrices is fixed with increasing angular order, and the typical expression of scattering kernels as Legendre moments makes the source/interaction terms easy to compute in many nuclear/radiative transfer applications. As such, the only two main barriers to scalable P$_n$ solutions in general Boltzmann transport problems are the calculation of half-range integrals/application of BCs across faces and the increasing conditioning problems when capturing discontinuities (e.g., problems with streaming) leading to an increase in iteration count.  It is these two ``scalability'' issues this paper aims to tackle. 

At best, the calculation of inflow/outflow across a face (or the application of BCs) with spherical harmonics is $\mathcal{O}(n^2)$ in angle size. There are two main approaches to computing these terms, including Riemann decomposition using an eigendecomposition \cite{Buchan2011}, or simply applying a quadrature to compute an approximate half-range integral. Both of these methods can benefit from the rotational invariance of spherical harmonics and the sparsity of the spherical harmonics angular matrix in the $z$ direction to reduce either the number of eigendecompositions calculated, or the number of quadrature points needed (for examples and reviews, please see \cite{Mohlenkamp1999, Park2006, Lessig2012}). Even if these calculations could be computed quickly, simply the application of the resulting dense Riemann decomposition on each face is $\mathcal{O}(n^2)$. The only feasible strategy (aside from a breakthrough in spherical harmonics rotation algorithms) to reducing the overall runtime is therefore to reduce the size of the angular expansions on each face if possible, through an angular adapt. 

The second ``scalability'' barrier is in problems with heavy transport, and the aforementioned filtered P$_n$ methods offers the hope of improved conditioning in streaming problems. \cite{Laboure2016b} (Fig. 3.5) for example shows the number of GMRES iterations required to solve a 2D duct problem with FP$_n$ as $n$ is increased (up to FP$_{17}$). They present an almost constant iteration count for certain filter values as the angular order is increased. This combined with the potential reduction in the size of high-order inflow/outflow calculations required due to angular adaptivity offers the potential for fast spherical harmonics calculations in general problems.
% ~~~~~~~~~~~~~
\section{Spatial discretisation}
\label{sec:sub-grid}
% ~~~~~~~~~~~~~
We use a sub-grid scale spatial discretisation in this work \cite{hughes_variational_1998, hughes_multiscale_2006, candy_subgrid_2008, buchan_inner-element_2010}, which provides stabilisation while allowing for low-memory use in general applications. A brief overview of the spatial discretisation is given below, for more details please see \cite{buchan_inner-element_2010}. We start by decomposing the solution to \eref{eq:bte} as $\psi = \phi + \theta$, where $\phi$ and $\theta$ are the solutions on the ``coarse'' and ``fine'' scales, respectively. We then represent each of the solutions with a different finite element representation (like \cite{hughes_multiscale_2006}); on the coarse scale we use a continuous representation spanned by $\eta_N$ basis functions, while on the fine scale we use a discontinuous representation spanned by $\eta_Q$ basis functions, or
\begin{equation}
\phi(\bm{r}, \bm{\Omega}) \approx \sum_{i=1}^{\eta_N} N_i(\bm{r}) \tilde{\phi}_i(\bm{\Omega}); \qquad \theta(\bm{r}, \bm{\Omega}) \approx \sum_{i=1}^{\eta_Q} Q_i(\bm{r}) \tilde{\theta}_i(\bm{\Omega}),
\label{eq:space}
\end{equation}
where $N_i$ and $Q_i$ are the basis functions for the continuous and discontinuous spaces respectively, with $\tilde{\phi}_i$ and $\tilde{\theta}_i$ being the associated expansion coefficients. 

To discretise in angle, we use the P$_n$ and filtered P$_n$ expansions described in detail in \secref{sec:pn}, though here we refer to an arbitrary angular discretisation. We can represent the expansion coefficients $\tilde{\phi}_i$ and $\tilde{\theta}_i$ in \eref{eq:space} with an arbitrary angular discretisation, with the basis functions $G_j(\bm{\Omega})$ and different numbers of basis functions $\eta_A^i$ and $\eta_D^i$ for each expansion coefficient in space, on the coarse and fine scales respectively. Finally if we write the expansion coefficients $\tilde{\phi}_{i,j}$ and $\tilde{\theta}_{i,j}$, we have 
\begin{equation}
\tilde{\phi}_i(\bm{\Omega}) \approx \sum_{j=1}^{\eta_A^i} G_j(\bm{\Omega}) \tilde{\phi}_{i,j}; \qquad \tilde{\theta}_i \approx \sum_{j=1}^{\eta_D^i} G_j(\bm{\Omega}) \tilde{\theta}_{i,j}.
\label{eq:angle}
\end{equation}
If we apply the FEM as usual to \eref{eq:bte}, by integrating and applying Green's theorem, we can recover the linear system
\begin{equation}
\begin{bmatrix}
\mat{A} & \mat{B} \\
\mat{C} & \mat{D} \\
\end{bmatrix}
\begin{bmatrix}
\bm{\Phi} \\
\bm{\Theta} \\
\end{bmatrix}
=
\begin{bmatrix}
\mat{S}_{\bm{\Phi}} \\
\mat{S}_{\bm{\Theta}} \\
\end{bmatrix},
\label{eq:SGS_full}
\end{equation}
or equivalently
\begin{equation}
(\mat{A} - \mat{B} \mat{D}^{-1} \mat{C}) \tilde{\bm{\Phi}} = \mat{S}_{\bm{\Phi}} - \mat{B} \mat{D}^{-1} \mat{S}_{\bm{\Theta}}.
\label{eq:SGS}
\end{equation}
where $\mat{S}_{\bm{\Phi}}$ and $\mat{S}_{\bm{\Theta}}$ are the discretised source and $\tilde{\bm{\Phi}}$ and $\tilde{\bm{\Theta}}$ are vectors containing the coefficients of the coarse and fine discretised solutions, $\tilde{\phi}_{i,j}$ and $\tilde{\theta}_{i,j}$, respectively. Further details of $\mat{A}, \mat{B}, \mat{C}$ and $\mat{D}$ are provided in \cite{buchan_inner-element_2010}, though we should note that \eref{eq:SGS} has the same number of DOFs as the continuous problem; indeed $\mat{A}$ is simply the linear system that would result from discretising \eref{eq:bte} with continuous finite elements. Once we have solved \eref{eq:SGS}, we can reconstruct the fine solution with
\begin{equation}
\bm{\Theta} = \mat{D}^{-1} (\mat{S}_{\bm{\Theta}} - \mat{C} \bm{\Phi}),
\label{eq:theta}
\end{equation}
and then form our discrete solution $\bm{\Psi} = \bm{\Phi} + \bm{\Theta}$ (the coarse solution $\bm{\Phi}$ can easily be projected onto the fine space when performing this addition). 

When we have an adapted P$_n$/FP$_n$ simulation, we also follow \cite{Goffin2014, Goffin2015a} and remove the coupling in $\mat{A}, \mat{B}, \mat{C}$ and $\mat{D}$ in \eref{eq:SGS} (and \eref{eq:theta}) between the angular moments at a node and any higher order moments on neighbouring nodes. Without this modification, non-physical discontinuities appear in the scalar flux; see \cite{Goffin2015a} for a more detailed explanation of why this is necessary.

The $\mat{B} \mat{D}^{-1} \mat{C}$ term in \eref{eq:SGS} can be considered as a stabilisation term. To improve the performance of our discretisation, we have made a number of approximations to $\mat{D}$ in order to decrease the cost of inverting it (see \cite{buchan_inner-element_2010, Dargaville_2014, Adigun2018}). These approximations do not affect the conservation of our scheme, as our conservation statement is on the coarse scale of our discretisation.

We begin (as in \cite{buchan_inner-element_2010, Dargaville_2014, Buchan2016, Dargaville2019}) by making $\mat{D}$ element local, by replacing the DG jump terms with a vacuum condition on all element boundaries. Furthermore, we enforce a block-diagonal form, but note that the block diagonal form is specific to the angular discretisation used. As such we use a different blocking to \cite{Dargaville2019}, instead of grouping over the spatial nodes in an element for each angular coefficient, we block together ``shells'' of angular coefficients across the element. In three dimensions, these shells correspond to four angular coefficients, in two dimensions either three or four angular coefficients are blocked together. Making this block size constant ensures that computing/storing our block diagonal $\mat{D}^{-1}$ remains scalable as the angular order increases. This block diagonal form is more expensive to form/store than that shown in \cite{Dargaville2019}, but produces stable results when used with P$_n$/FP$_n$. For example, if we have an element with three spatial nodes in two dimensions, with angular orders P$_1$, P$_3$ and P$_5$ on each of the nodes respectively, we have a maximum number of 21 angular basis functions present on each spatial node of the element. We group together the angular basis functions into the shells $[1-3,4-7,8-10,11-14,15-18,19-21]$ and remove the off-diagonal coupling in $\mat{D}$ between shells. The first shell consisting of basis functions 1 to 3 is present on all spatial nodes, hence that block is of size $9 \times 9$. The final shell of functions 19-21 however is only present on a single spatial node, giving a block size of $3 \times 3$.

Given the large cost of computing the inflow/outflow across a face discussed in \secref{sec:aadapt}, we further approximate $\mat{D}$ on internal faces by grouping together face normals across the spatial mesh that are ``close'' to each other. We consider a base normal $\bm{n}_1$ on a face and group other normals, say $\bm{n}_2$, with this base normal if in two dimensions $\bm{n}_1 \cdot \bm{n}_2 < \cos(\pi/500)$, and in three dimensions if $\bm{n}_1 \cdot \bm{n}_2 < 1 - \pi/10000$. We never apply this grouping on external faces, to ensure we always accurately represent boundary conditions on the domain of the problem. Importantly, if we have adapted angle, the spatial nodes present on a face may not have the same angular resolution (this is somewhat ameliorated by the resolution smoothing discussed in \secref{sec:Adaptivity algorithm}). This is further exacerbated when we group our internal face normals, as there may be many different angular orders that must be represented on a face across a single grouped normal. To calculate our inflow/outflow, we explicitly compute an eigendecomposition across the face (rather than using a quadrature method). We note of course that once we have computed our eigendecomposition, they are no longer hierarchical (that is the Riemann computation across a face at P$_{11}$ is not a subset of that at P$_{13}$). For large angular order however, they are close and so we simply calculate the Riemann decomposition for the highest angular order present on a grouped face. Given this is not a good approximation at low order, we also chose to group together normals that only include P$_1$/FP$_1$ approximations separately. We could extend this to only grouping together normals that are ``close'' to each other in angular size as well as normal direction, but we found that this does not give substantial benefit. 

Given the block-diagonal form of $\mat{D}$ described above, we need only store an block-diagonal form of the internal Riemann decompositions across our grouped normals. This storage (as well as that for the block-diagonal $\mat{D}^{-1}$) scales linearly with our adapted angle size, meaning we need only compute these decompositions once per adapt step and store the result, as we need to reuse them when computing our goal-based residuals (see \secref{sec:Goal-based adaptivity}). Furthermore, given the rotational invariance of the spherical harmonics expansion, we only need to compute an eigendecomposition once per grouped normal to calculate the Riemann decomposition needed for both the inflow and outflow across a face. We can also reuse both these decompositions in the adjoint problem if we are using goal-based error metrics. 

These features and modifications to $\mat{D}$ combined with our angular adaptivity significantly decreases both the number and size of eigendecompositions required by our discretisation, and result in a performant spherical harmonics algorithm. We discuss this further in \secref{sec:Results}.

We also scale our element blocks of $\mat{D}^{-1}$ by $\gamma$ ($0 < \gamma < 1$) defined in \cite{Ragusa2012}, to prevent locking in pure scatter regions. The benefit to solving \eref{eq:SGS} as opposed to standard DG formulation is that the static condensation (given the approximations applied to $\mat{D}$, our discretisation can be considered as formed from an approximate Schur-complement) allows us to solve for $\tilde{\bm{\Phi}}$ and then reconstruct $\tilde{\bm{\Psi}}$. Particularly in 3D, the size of $\tilde{\bm{\Phi}}$ on the CG mesh is much smaller than $\tilde{\bm{\Psi}}$ on the DG mesh. Furthermore, we use linear basis functions in both the continuous and discontinuous spatial expansions given by \eref{eq:space}, hence we can often reuse temporary data during our matrix-free matrix-vector product, making our matvecs less expensive in practice.
% ~~~~~~~~~~~~~
\section{Angular adaptivity}
\label{sec:Angular adaptivity}
% ~~~~~~~~~~~~~
% ~~~~~~~~~~~~~
\subsection{Error metrics}
\label{sec:Error metrics}
% ~~~~~~~~~~~~~
We consider two forms of angular adaptivity in this work, regular and goal-based adaptivity. We will refer to \eref{eq:bte} as the ``forward'' problem, with exact solution $\bm{\psi}_{\textrm{exact}}$ and residual $\mathcal{R}$, hence $\mathcal{R}(\bm{\psi}_{\textrm{exact}}) = 0$. In this section, we are trying to compute an approximation, $\mat{e}$, to the exact error, $\bm{\epsilon} = \bm{\psi}_{\textrm{exact}} - \bm{\psi}$, in order to guide our adaptivity.
% ~~~~~~~~~~~~~
\subsubsection{Regular adaptivity}
\label{sec:Regular adaptivity}
% ~~~~~~~~~~~~~
Regular adaptivity is simple with a hierarchical discretisation, as the coefficients in the expansion can be thresholded, with small coefficient guaranteeing small contribution to the norm of the function we are representing; our only job is picking a thresholding tolerance, $\tau$. We therefore define our regular error metric as $\mat{e} = \bm{\epsilon} \approx |\bm{\psi}|/\tau$. For convenience, in the results presented below, we also scale $\mat{e}$ by the maximum scalar flux across the problem; this simply helps make the choice of $\tau$ more problem agnostic. 
% ~~~~~~~~~~~~~
\subsubsection{Goal-based adaptivity}
\label{sec:Goal-based adaptivity}
% ~~~~~~~~~~~~~
Goal-based adaptivity focuses resolution wherever needed to reduce the error in some arbritrary functional. In this section, we briefly review the formulation of goal-based error metrics through a dual-weighted residual method, described by \cite{Goffin2015, Goffin2015a} and used by \cite{Dargaville2019}. We can write the goal of the calculation in terms of a functional, $F$, of the solution as
\[
F(\bm{\psi}) = \int_P f(\bm{\psi}) \, \textrm{d}P,
\]
where $f$ is an arbritrary function of the solution and $P$ represents the phase-space. Functionals can be easily defined for quantities such as the average flux over a region, current over given surfaces, reaction rates and even eigenvalues. We can approximate the error in our functional as
\begin{equation}
|F(\bm{\psi}_{\textrm{exact}}) - F(\bm{\psi})| \approx \bm{\epsilon}^{\textrm{T}} \mat{R}^*
\label{eq:error}
\end{equation}
or equivalently
\begin{equation}
|F(\bm{\psi}_{\textrm{exact}}) - F(\bm{\psi})| \approx \bm{\epsilon}^{*\textrm{T}} \mat{R}
\label{eq:adjoint_err}
\end{equation}
where $\bm{\epsilon}^{\textrm{T}}$ and $\bm{\epsilon}^{*\textrm{T}}$ are the discrete forward and adjoint solution error, respectively, with $\mat{R}$ and $\mat{R}^*$ the discrete forward and adjoint residuals computed using $\bm{\psi}^*$ and $\bm{\psi}^*_{\textrm{exact}}$, which are the approximate and exact solutions of the adjoint equation with source term derived from the response function respectively.

In order to avoid \eref{eq:error} and \eref{eq:adjoint_err} both being zero due to Galerkin orthogonality, we further approximate \eref{eq:error} and \eref{eq:adjoint_err} by computing ``reduced-accuracy'' discrete residuals $\hat{\mat{R}}$ and $\hat{\mat{R}}^*$. We must also pick a target error for our goal-based adaptivity, similar to the thresholding tolerance in Section \ref{sec:Regular adaptivity}; we denote this tolerance again as $\tau$. We form our approximate error metric for each angular coefficient by computing the pointwise maximum of both the forward and adjoint pointwise errors, and scaling by the target error in each DOF, namely
\begin{equation}
\mat{e} = \frac{\max\{|\bm{\epsilon} \odot \hat{\mat{R}}^*|, |\bm{\epsilon}^* \odot \hat{\mat{R}}|\} N_{\textrm{DOF}}}{\tau}, 
\label{eq:gb_metric}
\end{equation}
where $\odot$ denotes pointwise multiplication. The use of the $\max$ operator in \eref{eq:gb_metric} ensures that features present in both the forward and adjoint solutions are resolved by the adaptivity (we define the orientation of our adjoint angular space to be the opposite of the forward, so we can easily compute products involving both our forward and adjoint wavelet coefficients). We are now left to define both the solution errors $\bm{\epsilon}$ and $\bm{\epsilon}^*$ and the reduced-accuracy residuals.

Similar to the regular adaptivity, we choose $\bm{\epsilon} \approx |\bm{\psi}|$ and $\bm{\epsilon}^* \approx |\bm{\psi}^*|$. Rather than simply using a diagonal approximation for the reduced-accuracy residuals like \cite{Dargaville2019}, which lead to a pathological effectivity index, we compute our reduced accuracy coarse and fine residuals, $\hat{\mat{R}}_{\bm{\Phi}}$ and $\hat{\mat{R}}_{\bm{\Theta}}$, respectively, using
\begin{equation}
\begin{bmatrix}
\hat{\mat{R}}_{\bm{\Phi}} \\
\hat{\mat{R}}_{\bm{\Theta}} \\
\end{bmatrix}
=
\begin{bmatrix}
\tilde{\mat{A}} & \tilde{\mat{B}} \\
\tilde{\mat{C}} & \tilde{\mat{D}} \\
\end{bmatrix}
\begin{bmatrix}
\bm{\Phi}\\
\bm{\Theta} \\
\end{bmatrix} - 
\begin{bmatrix}
\mat{S}_{\bm{\Phi}} \\
\mat{S}_{\bm{\Theta}} \\
\end{bmatrix}.
\label{eq:disc_resid_subgrid}
\end{equation}
The modified submatrices, $\tilde{\mat{A}}, \tilde{\mat{B}}, \tilde{\mat{C}}$ and $\tilde{\mat{D}}$ (note we still apply the approximations described in \secref{sec:sub-grid} to $\tilde{\mat{D}}$) are formed by coarsening the angular resolution (reducing the angular order by 2) in the diagonal blocks corresponding to the multiplication of each spatial basis function by itself. Note this is not the same as forming a residual by coarsening both $\bm{\Phi}$ and $\bm{\Theta}$. These coarsened residuals can easily be calculated by a single matrix-free matrix-vector product, and will not be zero. These residuals are then combined like the discrete solution in \secref{sec:sub-grid} to form our reduced accuracy discrete residual, $\hat{\mat{R}}$, as
\begin{equation}
\hat{\mat{R}} = \hat{\mat{R}}_{\bm{\Phi}} + \hat{\mat{R}}_{\bm{\Theta}}  
\label{eq:disc_resid}
\end{equation}

We must also take care when using goal-based adaptivity in streaming regions, to ensure that the coarsest angular resolution used produces a response in our goal. If we do not, then the adaptivity algorithm will not refine. This is a problem faced by all goal-based error metrics in the presence of advection, not just in Boltzmann transport problems. In particular, we discussed this previous in \cite{Dargaville2019}, where ray-effects caused by a non-rotationally invariant angular discretisation can cause both the forward/adjoint solutions/residuals to be zero in areas that a coarse discretisation cannot ``see''. Our P$_n$/FP$_n$ discretisations are rotationally invariant so there are no ray-effects, but insufficient angular resolution can change how far radiation propogates along a streaming path. \secref{sec:Results} examines this further, but we note in general problems that P$_1$ may not be suitable as the coarsest angular discretisation used as part of a goal-based adaptivity simulation.
% ~~~~~~~~~~~~~
\subsection{Adaptivity algorithm}
\label{sec:Adaptivity algorithm}
% ~~~~~~~~~~~~~
We now consider our iterative algorithm for the angular adaptivity. We begin the first adapt step by first solving the forward linear system  with our coarse angular discretisation, then solve the coarse adjoint linear system if goal-based adaptivity is used. We then compute the regular/goal-based error metric and perform refinement/coarsening. This is then followed by further adapt steps, up to some maximum refinement level.

As mentioned, the direction space $\Omega$ in our adjoint problem is explicitly written as the negative of our forward angular domain (i.e., our adjoint angular domain is a reflection about the origin). Our error metric \eref{eq:gb_metric} ensures that refinement is triggered in areas important to both the forward and adjoint solutions. We do this as it simplifies our implementation, as we can then apply the ``same'' angular discretisation to our forward and adjoint problems. A disadvantage of this approach is we may be applying more DOFs in angle than by performing adaptivity separately for the forward and adjoint problems. In practice however, we find this is not significant.

Given our spatial discretisation described in \secref{sec:sub-grid}, the error metrics given in \secref{sec:Error metrics} are all computed using $\bm{\Psi}$, which is formed from the sum of our coarse and fine solutions. This solution and hence our error metrics are computed on the fine mesh (i.e., the DG mesh), but we perform our angular adaptivity on the CG mesh. We therefore take the maximum error over the DG nodes that share their position with each CG node, to form an error metric, $\tilde{\mat{e}}$, on the CG mesh for each angular coefficient. We then take the maximum of the angular coefficients in $\tilde{\mat{e}}$ over each CG node. This gives us a single coefficient per CG spatial node $i$, which we denote as $\tilde{\mat{e}}_i^{\textrm{max}}$, which drives our adaptivity.  At each CG spatial node, we trigger refinement and increase the order of our expansion by two if $\tilde{\mat{e}}_i^{\textrm{max}} \geq 1.0$. We allow the order to increase by four if $\tilde{\mat{e}}_i^{\textrm{max}} \geq 2.0$, and allow coarsening by reducing the order by two if $\tilde{\mat{e}}_i^{\textrm{max}} < 0.1$.

We then smooth the resulting angular orders at each CG spatial node by averaging the angular order across the mesh connectivity of that CG spatial node. This ensures we have a smooth transition between areas of low and high angular resolution. The expansion order present on the DG nodes of the spatial mesh are then slaved to their CG counterparts and share the same angular discretisation. The choice to adapt on the CG spatial mesh means that adjacent faces in our mesh share the same angular discretisation. In intermediate adapt steps, to improve our runtimes we reduce the tolerance of our linear solves, as only the final linear solve with the finest discretisation needs to be solved to a high tolerance. For all adapted P$_n$/FP$_n$ simulations, the linear solves prior to the final step are to a relative tolerance of 1\xten{-4}, with the final step solved to 1\xten{-10}.
% ~~~~~~~~~~~~~
\section{Spatially dependent filter}
\label{sec:Spatially dependent filter}
% ~~~~~~~~~~~~~
As mentioned in \secref{sec:fpn}, we can allow the filter strength, $\Sigma_{\textrm{f}}$, to be spatially dependent. \cite{Laboure2016} suggest two different ways to set the value of $\Sigma_{\textrm{f}}$. The first of these is running a calculation with no filter, and then filtering where the scalar flux in the original problem is negative, with the second using a coarse spatial mesh and a high angular order ($N_0$) to determine an acceptable unfiltered solution, then computing the filter values as $\Sigma_{\textrm{f}} = \Sigma_{\textrm{t}}/f(1, N_0)$. Unfortunately neither of these approaches are suitable for problems with heavy streaming, as we found even (unfiltered) P$_1$ solutions in many problems are too poorly conditioned to solve. As such, we take a different approach and connect the value of our spatially dependent $\Sigma_{\textrm{f}}$ to our adaptivity algorithm discussed in \secref{sec:Adaptivity algorithm}.

We start by picking a constant (large) $\Sigma_{\textrm{f}}$ across the entire spatial domain for our first, coarse angular adapt step. This ensures that even the FP$_1$ solution can be easily computed in the first adapt step. After this first step, we note that we have computed the ``coarse'' and ``fine'' scale sub-grid components, $\bm{\Phi}$ and $\bm{\Theta}$, (and of course our full angular flux, $\bm{\Psi}$) due to the decomposition performed as part of our spatial discretisation (see \secref{sec:sub-grid}). The fine solution, $\bm{\Theta}$, represents the amount of stabilisation applied to each angular coefficient at each of the discontinuous spatial nodes. If the solution is smooth (in space/angle), $\bm{\Theta}$ is small, and if there is a significant discontinuity, $\bm{\Theta}$ is larger. The calculation of $\bm{\Theta}$ is also agnostic to the cause of the discontinuity, which is vital. 

Given this, it is then easy to determine a spatially dependent filter value by using $\bm{\Theta}$ to scale $\Sigma_{\textrm{f}}$ so it is large in spatial regions where discontinuities are present. We begin by computing $\Sigma_{\textrm{stab}}$, the integral of $\bm{\Theta}$ over angle at each of the discontinuous nodes,
\begin{equation}
\Sigma_{\textrm{stab}} = \int_{\bm{\Omega}} \bm{\Theta} \, \textrm{d}\bm{\Omega},
\label{eq:int_stab}
\end{equation}
which is trivial to compute given the first isotropic moment of a P$_n$/FP$_n$ expansion. We then take the maximum (in magnitude) of $\Sigma_{\textrm{stab}}$ across the DG nodes and form $\tilde{\Sigma}_{\textrm{stab}}$ on the CG nodes. If we denote the constant filter value we set in the first adapt step as $\Sigma_{\textrm{f}}^1$, subsequent spatially dependent filter values are then calculated as
\begin{equation}
\Sigma_{\textrm{f}} = \Sigma_{\textrm{f}}^1 \left(\frac{|\tilde{\Sigma}_{\textrm{stab}}|}{\max(|\tilde{\Sigma}_{\textrm{stab}}|)}\right)^{(1/3)}.
\label{eq:space_filter}
\end{equation}
Much in the same way as our adaptive angular order, the DG values of this spatially dependent $\Sigma_{\textrm{f}}$ are then simply taken from their equivalent CG values. Equation \ref{eq:space_filter} has the effect of scaling down $\Sigma_{\textrm{f}}^1$ across the spatial grid. The cube root we use in \eref{eq:space_filter} is not necessary, we simply use it to compress the possible filter values, as the magnitude of $\bm{\Omega}$ can vary by many orders of magnitude. We can also scale $\Sigma_{\textrm{stab}}$ by the scalar flux at each spatial node, to ensure the filter value remains consistent in problems where the angular flux varies by many orders of magnitudes (we do this in one problem in \secref{sec:Results}). As for chosing a value of $\Sigma_{\textrm{f}}^1$, experimentation suggests a filter value of $\Sigma_{\textrm{f}}^1=1$ is a good starting value in many problems, though \eref{eq:space_filter} often has the effect of reducing the importance of picking a ``good'' $\Sigma_{\textrm{f}}^1$ (we discuss this in \secref{sec:Results}).

Equation \ref{eq:space_filter} is trivial to compute and naturally means the filter strength changes as we progress through our adaptive process. This is particular important, as we find that on the boundaries of spatial regions where we have adapted (for both P$_n$ and constant filter strength FP$_n$), if we move from a first order expansion $N=1$ to any other angular order, we see visible discontinuities in the scalar flux, in some problems (see the discussion in \secref{sec:2D dogleg problem}). \cite{Goffin2015a, Goffin2015} also found this (e.g., see Figure 2 in \cite{Goffin2015}), but did not discuss it. We find that this does not seem to occur in the transition region between higher-order expansions. One simple remedy would be to have our coarsest angular discretisation be larger than $P_1$/FP$_1$ with a cost of reducing the effectivity of our angular adaptivity. Instead we chose to rely on our spatially dependent filter. As mentioned, the fine scale solution, $\bm{\Theta}$, stabilises discontinuities, regardless of the cause. As such, we find that the value of $\bm{\Theta}$ in subsequent adapt steps around these regions is large and so our spatially dependent filter is also large, smoothing out the discontinuities. This is a powerful feature of our adaptive filtered P$_n$ method. It is also easy to see how \eref{eq:space_filter} could also be scaled by the goal-based error metrics described in \secref{sec:Goal-based adaptivity}, allowing very high filter values in unimportant regions of the problem, improving the overall conditioning and decreasing the iteration count in the linear solver. We have found success in doing so, but discussing this in detail is left for future work.

When performing goal-based adaptivity, we also explicitly allow different spatially dependent filter values for our forward and adjoint problems. We could easily combine the values of $\Sigma_\textrm{stab}$ across both both the forward and adjoint problems, but we found in many problems that the difference in magnitude between the forward and adjoint solutions at the same spatial position meant both ended up overfiltered. Given these separate spatially dependent filter values for the forward/adjoint solutions, care should be taken when forming the error metrics \eref{eq:error}--\eref{eq:gb_metric}. For example in \eref{eq:error}, the discrete forward error and discrete adjoint residuals are convolved, though the angular coefficients at corresponding indices represent different filtered P$_n$ expansions. The solution for this would be to project onto a common space and then form the error metric; we however do not do this. For simplicity, we ignore the different filter values and calculate our error metrics as written in \secref{sec:Goal-based adaptivity}. We find this is an acceptable solution given our error metrics are at best an approximation; \secref{sec:2D dogleg problem} discusses the impact this has on our effectivity index.
% ~~~~~~~~~~~~~
\section{Linear solver}
\label{sec:Linear solver}
% ~~~~~~~~~~~~~
We use the same matrix-free method as \cite{Dargaville2019} to solve our linear system, namely FGMRES(30) preconditioned by a matrix-free multigrid method. Both the spatial tables and $\mat{D}^{-1}$ on the lower multigrid levels are built by using an agglomerate-local Galerkin projection on the top-grid element matrices. The lower-grid CG and DG spatial nodes take the adapted angles present on their fine equivalent, and the lower-grid CG and DG spatial nodes also use the filter value, $\Sigma_{\textrm{f}}$, present on their fine equivalent. Importantly, the fact that we do not use jump-terms in our discretisation means we do not need to worry about the non-straight element boundaries on the lower grids, formed from agglomerates of unstructured elements (or having to group these non-straight ``normals'' as described in \secref{sec:sub-grid}).  This allows us to perform matrix-free matrix-vector products on all multigrid levels. Our smoothers differ from \cite{Dargaville2019}, as the block-grouping of $\mat{D}^{-1}$ for P$_n$/FP$_n$ makes it difficult to assemble a diagonal scalably. As such, we use GMRES(3) preconditioned by Jacobi up to P$_{7}$/FP$_7$, then switch simply to GMRES(3) for higher order discretisations.
% ~~~~~~~~~~~~~
\section{Results}
\label{sec:Results}
% ~~~~~~~~~~~~~
We use three examples to test our angular adaptivity algorithm, with varying levels of smoothness. We expect the FP$_n$ algorithms presented to be of most benefit in problems with little smoothness. We start the adaptivity algorithm with a coarse uniform resolution of P$_1$/FP$_1$ unless otherwise stated and set a maximum level of refinement. Memory use is profiled using massif (from valgrind) which measures peak heap usage. Very little memory in our simulations is not on the heap, so this gives an accurate measurement of our total peak memory use. 

All non-P$_n$ solutions shown use are taken from \cite{Dargaville2019} for comparative purposes (using the same mesh, run on the same machine with the same compiler/optimisation flags, etc). These other angular discretisations, along with the adapted P$_n$ and FP$_n$ discretisations presented in this work have been implemented in FETCH2, the multi-physics, coupled Boltzmann transport code developed at the AMCG. Both the P$_n$ and FP$_n$ use high order P$_n$ calculations on the same spatial meshes as reference solutions, as we know the FP$_n$ converges to P$_n$ as $N \rightarrow \infty$.

We should note that for all the adapted simulations shown in this Section, the runtime shown includes all the adapt steps (i.e., all the linear solves performed, computation of error metrics, refinement/coarsening, etc) required to get to that order. For example, if we perform a regular adaptive simulation with 5 adapt steps, the runtime shown includes the time required to perform 5 linear solves. An equivalent goal-based simulation would include 9 linear solves; we do not solve the adjoint problem on the final adapt step. All uniform linear solves were performed in serial to an absolute/relative tolerance of 1\xten{-10}. 
% ~~~~~~~~~~~~~
\subsection{Brunner lattice problem}
\label{sec:Brunner lattice problem}
% ~~~~~~~~~~~~~
% ~~~~~~~~~~~~~
The first example is the lattice problem from \cite{Brunner2002}. We discretise this problem in space with the same mesh used by \cite{Dargaville2019}, namely an unstructured triangular mesh with 3378 elements (1690 CG nodes and 10,134 DG nodes). This problem has regions of smoothness in angle, but still features discontinuities, particularly in the corners between the scattering and absorbing regions. Previously, we showed \cite{Dargaville2019} that uniform P$_n$ performed well in this problem compared to other angular discretisations.  We use regular adaptivity in this problem, as large regions of the phase-space are important to the final solution and we allow a maximum of 10 adapt steps for both P$_n$ and FP$_n$. The reference solution used is uniform P$_{101}$ with 5253 DOFs in angle, using 60M DOFs. We compute the relative error in the 2-norm of the scalar flux in this problem.

\begin{figure}[th]
\centering
\subfloat[][Error vs CDOFs]{\label{fig:brunner_convg_pn}\includegraphics[width =0.47\textwidth]{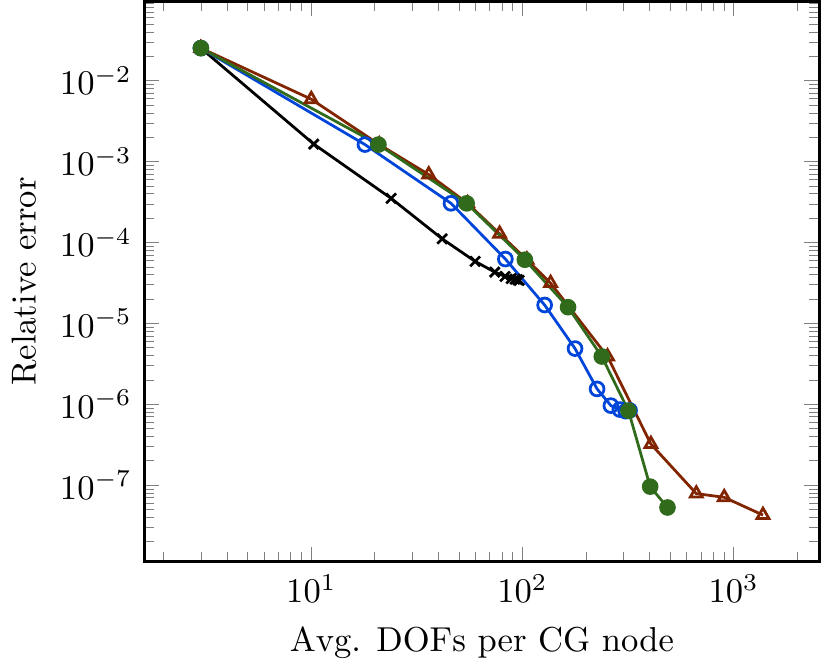}}
\subfloat[][Error vs total runtime]{\label{fig:brunner_time_pn}\includegraphics[width =0.47\textwidth]{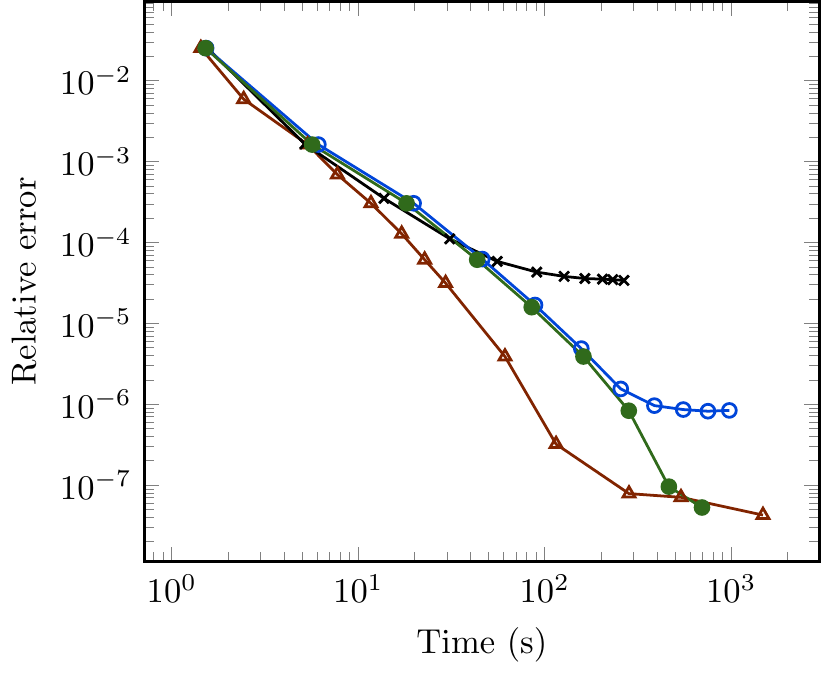}}\\
\caption{Performance of the regular angular adaptivity with P$_n$, in the relative error of the 2-norm of the scalar flux across the domain, for the Brunner problem. The x, \textcolor{matlabblue}{o} and \textcolor{foliagegreen}{\CIRCLE} markers use threshold coefficients 1\xten{-3}, 1\xten{-4} and 1\xten{-5}, respectively, with the \textcolor{fireenginered}{$\triangle$} uniform (unadapted).}
\label{fig:brunner_adapt}
\end{figure}

To begin, we examine the impact of the thresholding tolerance used; too small of a tolerance and the adapt process will add unecessary angles, too large and it will not add enough to reach a desired error. \fref{fig:brunner_adapt} shows the results from modifying the thresholding tolerance from 1\xten{-3} to 1\xten{-5}. We can see in \fref{fig:brunner_convg_pn} that a tolerance of 1\xten{-3} and 1\xten{-4} causes the adaptivity to plateu with the number of CDOFs, with 1\xten{-5} producing a solution that matches the uniform P$_n$ until we reach an order of P$_{21}$. At this point the adapted P$_n$ produces a solution of equivalent accuracy with less CDOFs than the uniform. \fref{fig:brunner_time_pn} shows that the uniform P$_n$ is quicker than the adapted P$_n$, until we reach uniform P$_{41}$ and P$_{51}$, when the adapted is quicker. This is because, for low order, we do not save a substantial number of DOFs by adapting in this problem, and the cumulative cost of solving the linear system at each adapt step outweights the cost of the Riemann decompositions. At higher order however, this balance changes. 
% ~~~~~~~~~~~~~
% ~~~~~~~~~~~~~

\begin{figure}[th]
\centering
\subfloat[][Error vs CDOFs]{\label{fig:brunner_convg_fpn}\includegraphics[width =0.47\textwidth]{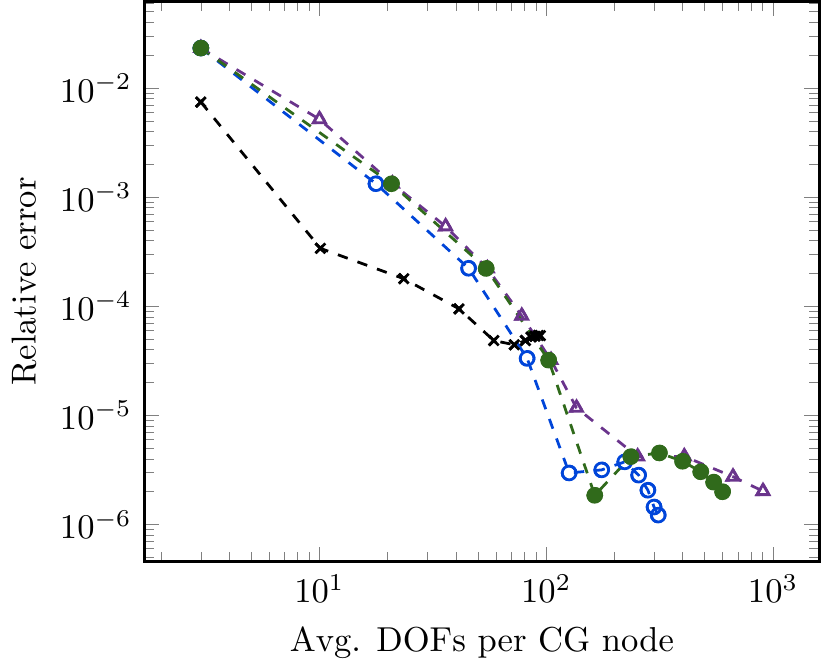}}
\subfloat[][Error vs total runtime]{\label{fig:brunner_time_fpn}\includegraphics[width =0.47\textwidth]{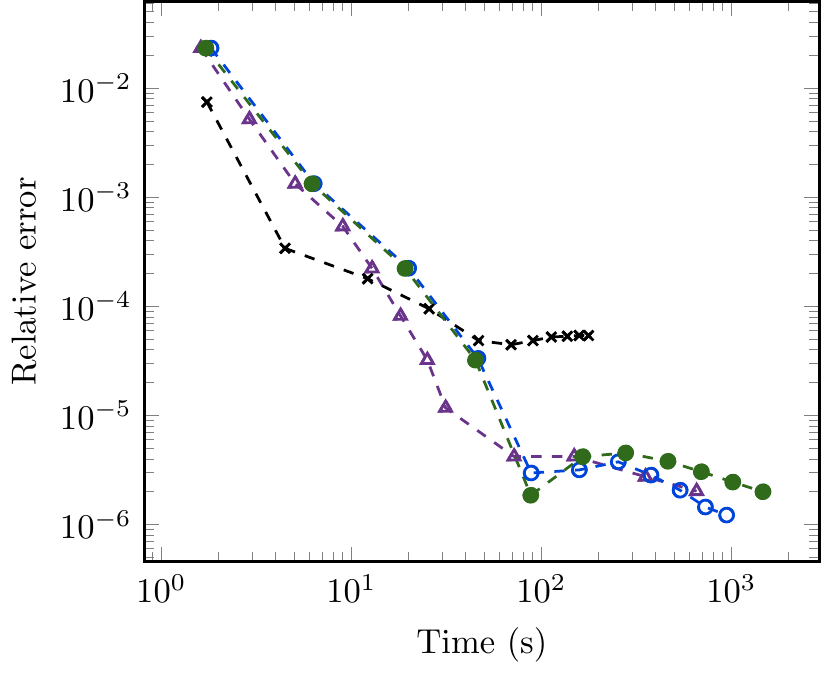}}\\
\caption{Performance of the regular angular adaptivity with FP$_n$ with $\Sigma_{\textrm{f}}=1$, in the relative error of the 2-norm of the scalar flux across the domain, for the Brunner problem. The x, \textcolor{matlabblue}{o} and \textcolor{foliagegreen}{\CIRCLE} markers use threshold coefficients 1\xten{-3}, 1\xten{-4} and 1\xten{-5}, respectively, with the \textcolor{gaylordpurple}{$\triangle$} uniform (unadapted).}
\label{fig:brunner_adapt_fpn}
\end{figure}

To begin examining the results from the FP$_n$ method, we fix the filter strength to a constant value and examine the impact of changing the thresholding tolerance, to verify that our FP$_n$ adaptivity still behaves similarly to the P$_n$. We do not focus heavily on adapted FP$_n$ with a constant filter strength in this paper, for reasons that will become evident below. \fref{fig:brunner_adapt_fpn} shows the impact of setting the filter strength to a constant $\Sigma_{\textrm{f}}=1$. We can see in \fref{fig:brunner_convg_fpn} that even with uniform angular resolution, the filter has significantly degraded the convergence in this problem compared with \fref{fig:brunner_convg_pn}, to the point of non-monotonicity. This is to be expected, as this problem has enough smoothness to benefit from the spectral nature of P$_n$. Examining the impact of the adaptivity and the thresholding tolerance, we see that again, choosing a tolerance that is too large (1\xten{-3}) causes the adaptivity to plateu. We can also see in \fref{fig:brunner_convg_fpn} that choosing a tolerance that is too small (1\xten{-5}) causes the adaptivity to include too many angles, whereas a tolerance of 1\xten{-4} achieves both a reduction in DOFs for a given error and a decrease in runtime, shown in \fref{fig:brunner_time_fpn}.
% ~~~~~~~~~~~~~
% ~~~~~~~~~~~~~

\begin{figure}[th]
\centering
\subfloat[][P$_n$]{\label{fig:brunner_pn_bottom_left_rotate}\includegraphics[width =0.4\textwidth]{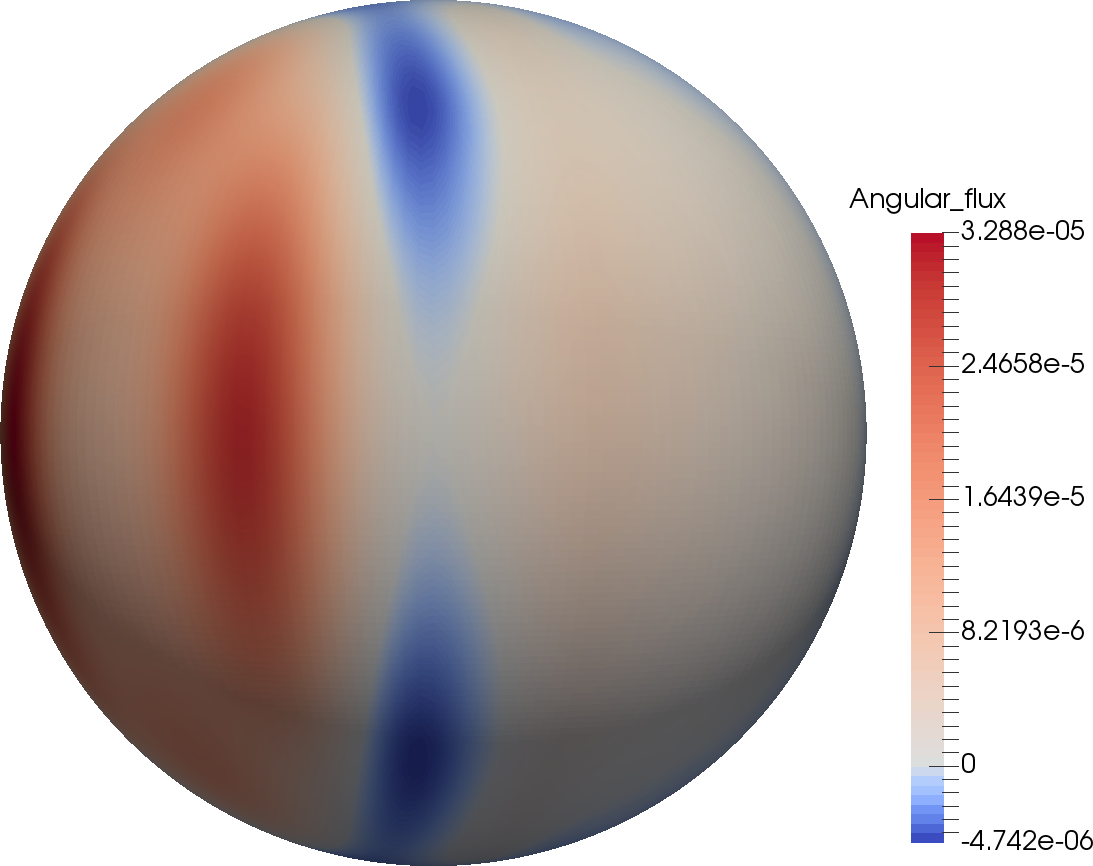}} \hspace{0.5cm}
\subfloat[][FP$_n$ with $\Sigma_{\textrm{f}}=100$]{\label{fig:brunner_fpn_filter_100_bottom_left_rotate}\includegraphics[width =0.4\textwidth]{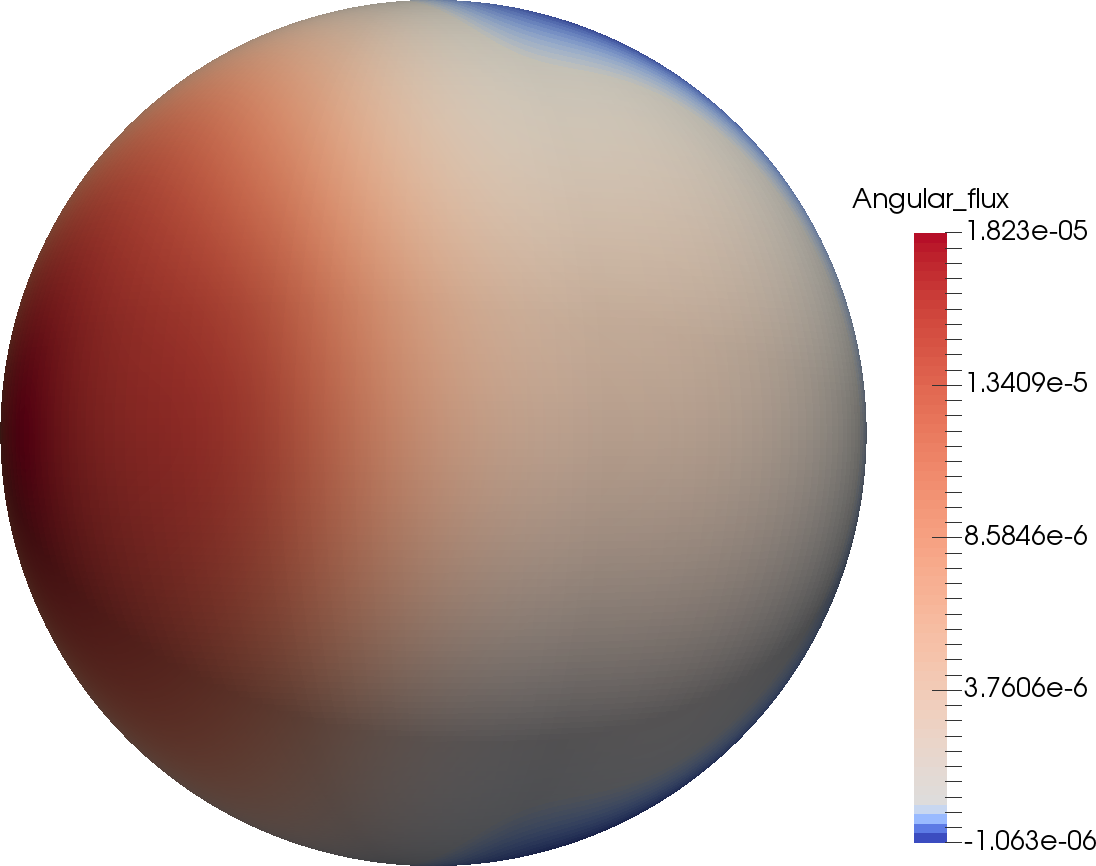}}\\
\caption{Angular flux in the Brunner problem at $x=2.5, y=2.5$, with regular angular adaptivity after 5 adapt steps, with threshold coefficient 1\xten{-3}.}
\label{fig:brunner_filter}
\end{figure}

\fref{fig:brunner_filter} visualises the impact of using filtered P$_n$ on the angular flux at a single point in the Brunner problem. \fref{fig:brunner_pn_bottom_left_rotate} shows that the P$_n$ solution features oscillations across the sphere, with a strong filter of $\Sigma_{\textrm{f}}=100$ smoothing the angular flux considerably, and decreasing the magnitude of negativity as expected.
% ~~~~~~~~~~~~~
% ~~~~~~~~~~~~~

\begin{figure}[th]
\centering
\subfloat[][P$_n$ - 960K DOFs]{\label{fig:brunner_pn_no_angles_7_steps_1e-3}\includegraphics[width =0.4\textwidth]{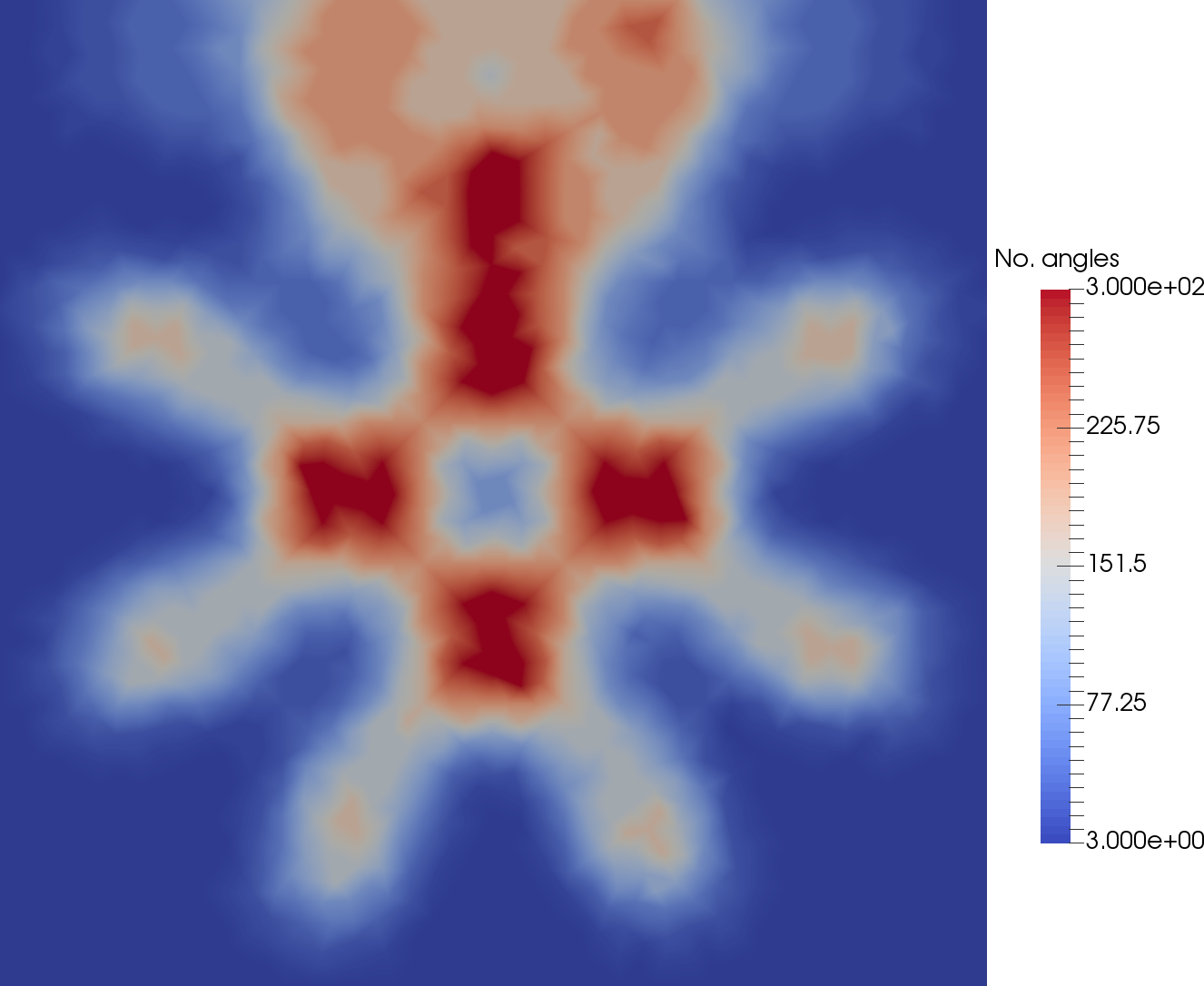}} \hspace{0.5cm}
\subfloat[][FP$_n$ with $\Sigma_{\textrm{f}}=100$ - 756K DOFs]{\label{fig:brunner_fpn_no_angles_7_steps_1e-3}\includegraphics[width =0.4\textwidth]{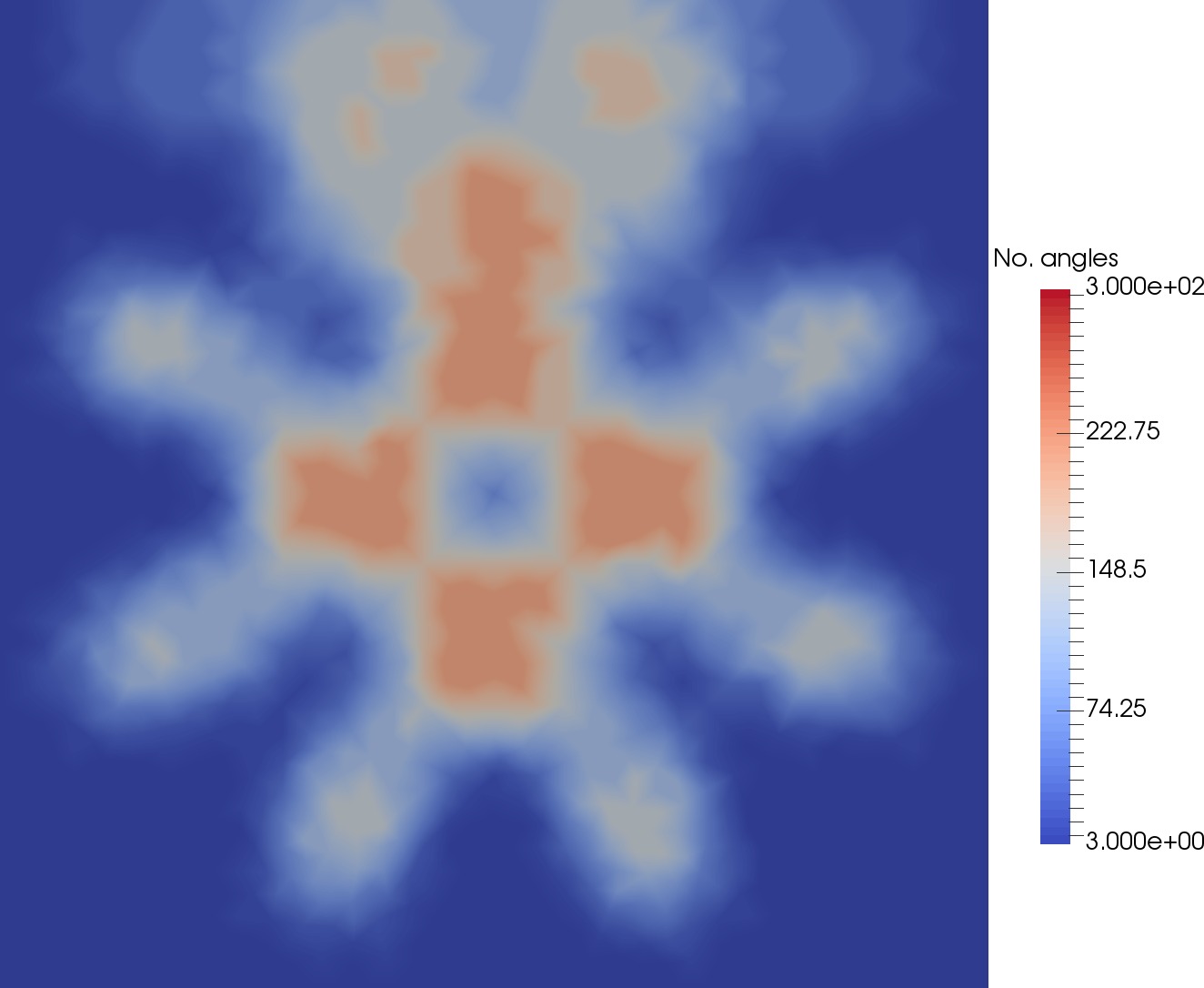}}\\
\caption{Number of basis functions across the spatial domain for the Brunner problem, plotted on the CG mesh, on the 7th step of regular angular adaptivity with threshold coefficient 1\xten{-3}.}
\label{fig:brunner_no_angles}
\end{figure}

\fref{fig:brunner_no_angles} shows where the adaptivity for both the P$_n$ and FP$_n$ with a strong constant filter has chosen to place angular resolution, and we can see that as expected when using regular adaptivity, the areas of high resolution are focused on areas with large flux. We can see in \fref{fig:brunner_fpn_no_angles_7_steps_1e-3} that the FP$_n$ has used fewer angles than the P$_n$ shown in \fref{fig:brunner_pn_no_angles_7_steps_1e-3} for a given tolerance (this is dependent on the filter strength/type and problem). 
% ~~~~~~~~~~~~~
% ~~~~~~~~~~~~~

\begin{figure}[th]
\centering
\subfloat[][Error vs CDOFs]{\label{fig:brunner_convg_fpn_space}\includegraphics[width =0.47\textwidth]{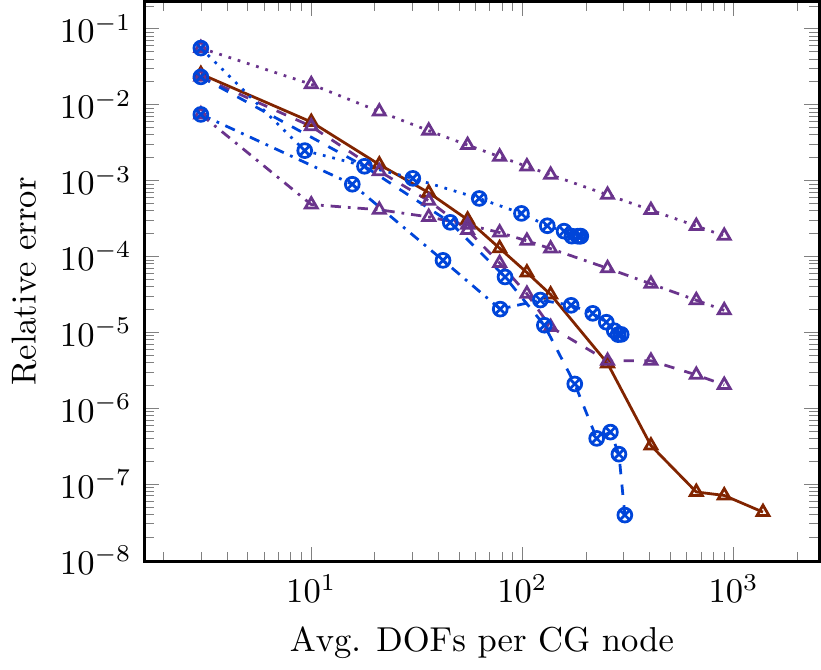}}
\subfloat[][Error vs total runtime]{\label{fig:brunner_time_fpn_space}\includegraphics[width =0.47\textwidth]{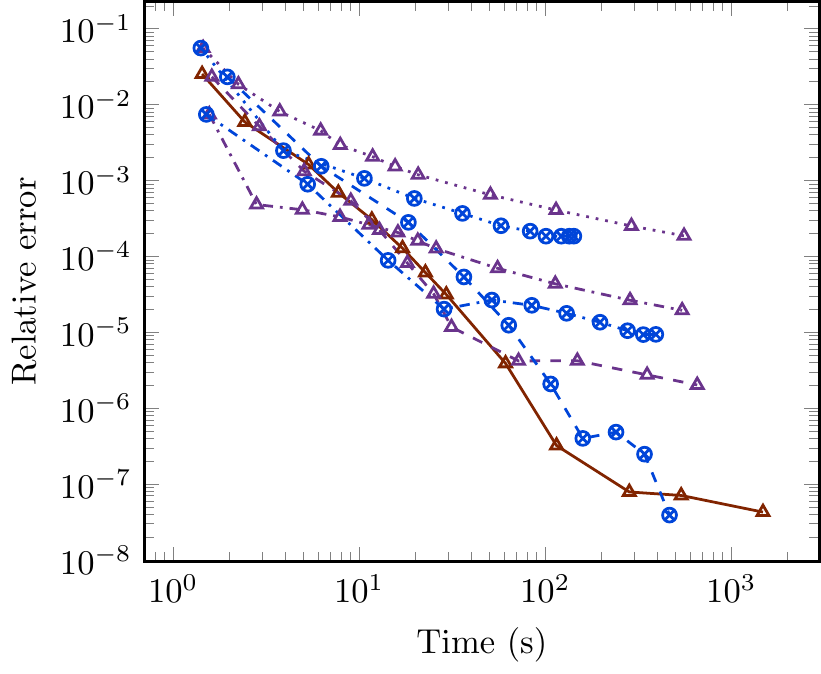}}\\
\caption{Investigating the impact of a spatially-dependent filter strength with FP$_n$, in the relative error of the 2-norm of the scalar flux across the domain, for the Brunner problem. The solid \textcolor{fireenginered}{$\triangle$} is uniform P$_n$, with the dotted \textcolor{gaylordpurple}{$\triangle$} FP$_n$ with $\Sigma_{\textrm{f}}=100$, dash-dotted FP$_n$ with $\Sigma_{\textrm{f}}=10$ and dashed FP$_n$ with $\Sigma_{\textrm{f}}=1$. The \textcolor{matlabblue}{$\otimes$} are regular adapted FP$_n$ with threshold tolerance 1\xten{-4}, with spatially dependent filter strength and reduced tolerance solves, with the dotted $\Sigma_{\textrm{f}}^{\textrm{1}}=100$, the dash-dotted $\Sigma_{\textrm{f}}^{\textrm{1}}=10$ and dashed $\Sigma_{\textrm{f}}^{\textrm{1}}=1$}
\label{fig:brunner_adapt_fpn_space}
\end{figure}

We can now turn to examining the behaviour of our FP$_n$ method as we change the filter strength. \fref{fig:brunner_filter} showed that a filter degrades convergence in a global norm in this problem, as theoretical results predict \cite{Frank2016}.  \fref{fig:brunner_adapt_fpn_space} shows the results from using both uniform FP$_n$ with a fixed constant filter strength, and comparing to an adapted FP$_n$ simulation with spatially dependent filter values. If we examine \fref{fig:brunner_convg_fpn_space} we can see that the uniform FP$_n$ all converge worse than the uniform P$_n$ in this problem. As we decrease the size of $\Sigma_\textrm{f}$ from 100 to 1, the convergence improves, though like in \fref{fig:brunner_filter} we see non-monotonic convergence. \fref{fig:brunner_time_fpn_space} also shows starts to show the effect of computing the Riemann decompositions in this problem at high order, as the runtimes of the uniform FP$_n$ method are increasing nonlinearly with the NDOFs. 

When we enable our spatially dependent filter strength, we can see in \fref{fig:brunner_convg_fpn_space} that for the same initial filter strength $\Sigma_\textrm{f}^1$ of between 100 and 1, the convergence is improved significantly compared to when the same constant filter strength is applied. Furthermore, the adapted FP$_n$ method with $\Sigma_\textrm{f}^1=1$ is significantly outperforming the uniform P$_n$ method per DOF, reaching an equivalent accuracy at high order with an average of 305 DOFs per CG node, compared to 1378 DOFs. We can see in \fref{fig:brunner_time_fpn_space} that this adapted calculation at high order is also approximately three times faster to compute. 
% ~~~~~~~~~~~~~
% ~~~~~~~~~~~~~

\begin{figure}[th]
\centering
\subfloat[][Absolute value of $\bm{\Sigma}_{\textrm{stab}}$ across space. This can be considered the net ``amount'' of stabilisation we apply at each node.]{\label{fig:brunner_fpn_stab_10_steps_1e-4_space_red}\includegraphics[width =0.47\textwidth]{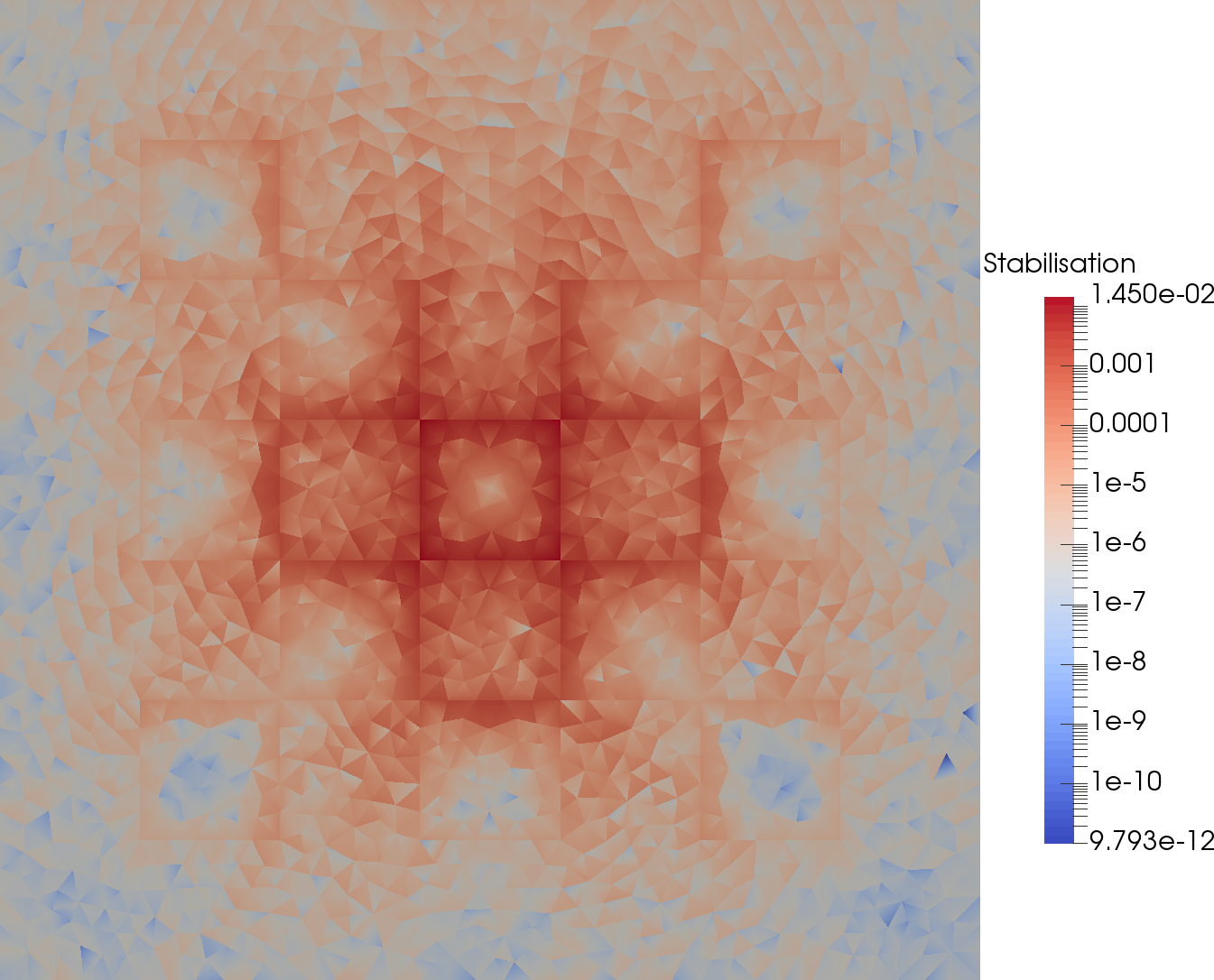}}
\subfloat[][Spatially-dependent $\Sigma_{\textrm{f}}$]{\label{fig:brunner_fpn_filter_10_steps_1e-4_space_red}\includegraphics[width =0.47\textwidth]{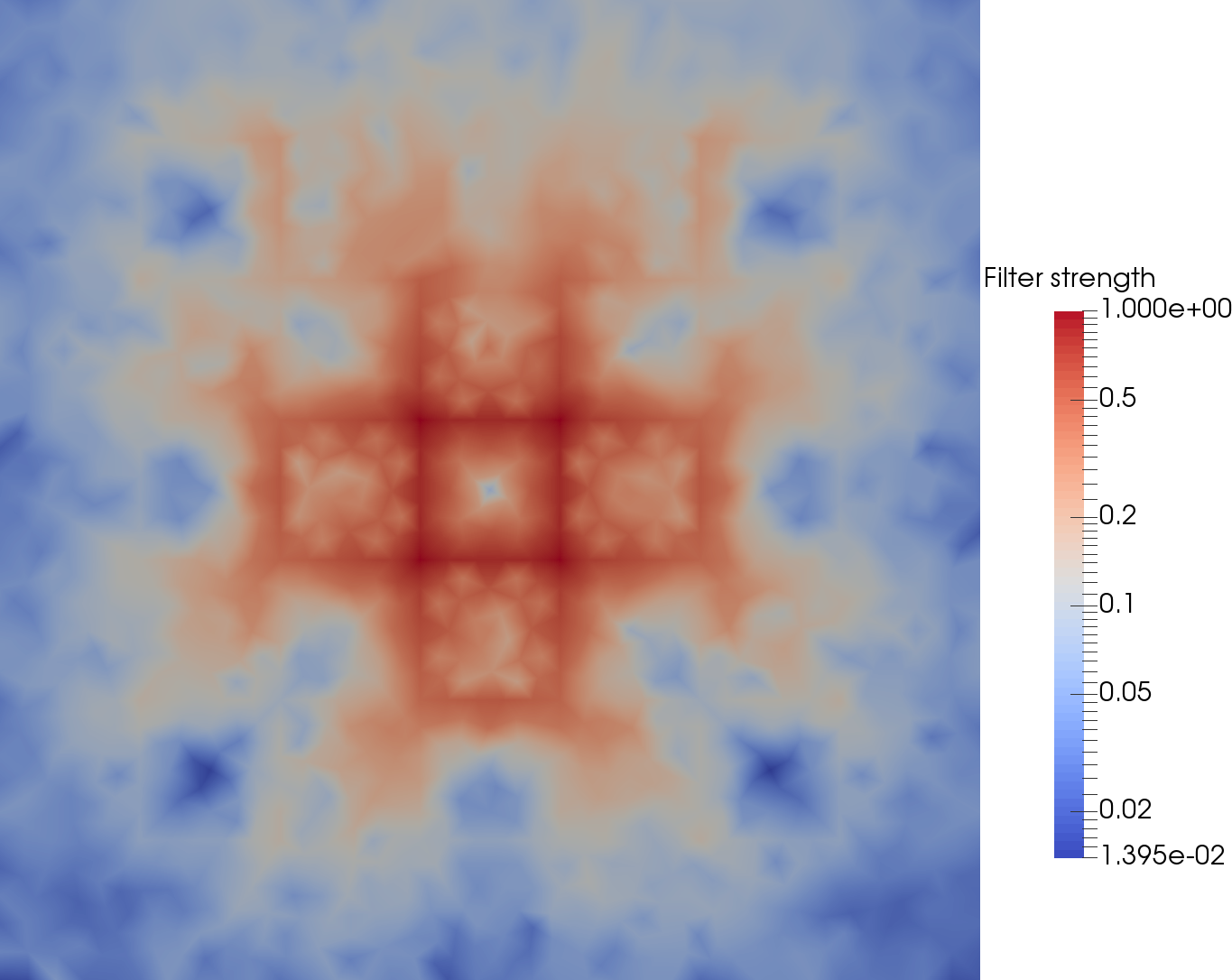}}\\
\caption{Computation of a spatially dependent $\Sigma_{\textrm{f}}$, with $\Sigma_{\textrm{f}}^{\textrm{1}}=1$, on the the 10th step of FP$_n$ regular adaptivity with threshold coefficient 1\xten{-4} for the Brunner problem.}
\label{fig:brunner_space}
\end{figure}

\fref{fig:brunner_space} shows the size of both $\bm{\Sigma}_\textrm{stab}$ and our spatially dependent $\Sigma_\textrm{f}$ in this adapted FP$_n$ simulation. We can see in \fref{fig:brunner_fpn_stab_10_steps_1e-4_space_red} that our net stabilisation is applied heavily around the material property boundaries in this problem, in greatest magnitude near the source. This is where we would expect discontinuities in space/angle to be large in this problem. \fref{fig:brunner_fpn_filter_10_steps_1e-4_space_red} shows the resulting filter values applied in this problem, where the largest value is given by $\Sigma_\textrm{f}^1=1$, and scaled down elsewhere in space according to \eref{eq:space_filter}. This results in a significantly smaller average filter value applied in this problem, and as noted improves convergence. 
% ~~~~~~~~~~~~~
% ~~~~~~~~~~~~~

\begin{figure}[th]
\centering
\subfloat[][Error vs CDOFs]{\label{fig:brunner_convg}\includegraphics[width =0.47\textwidth]{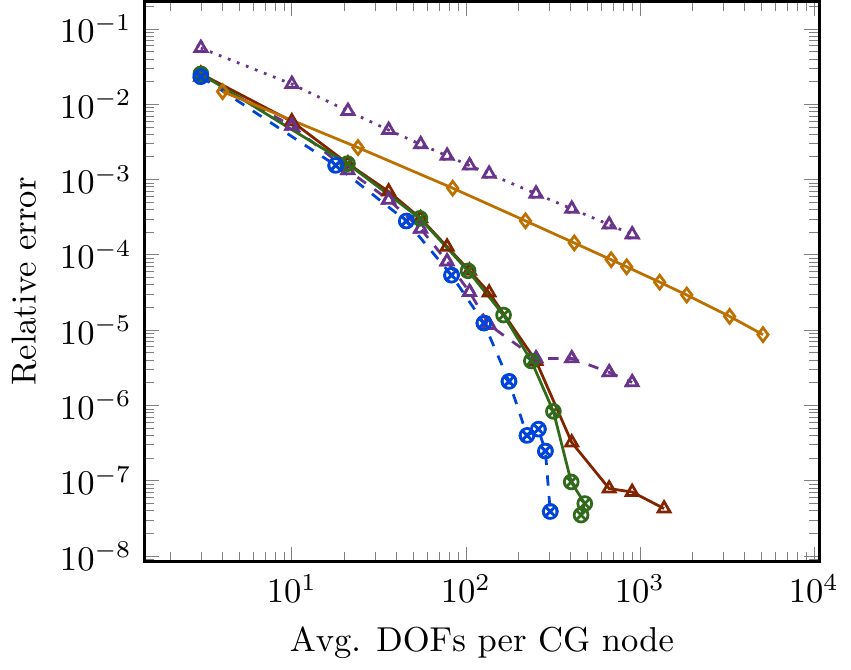}}
\subfloat[][Error vs total runtime]{\label{fig:brunner_time}\includegraphics[width =0.47\textwidth]{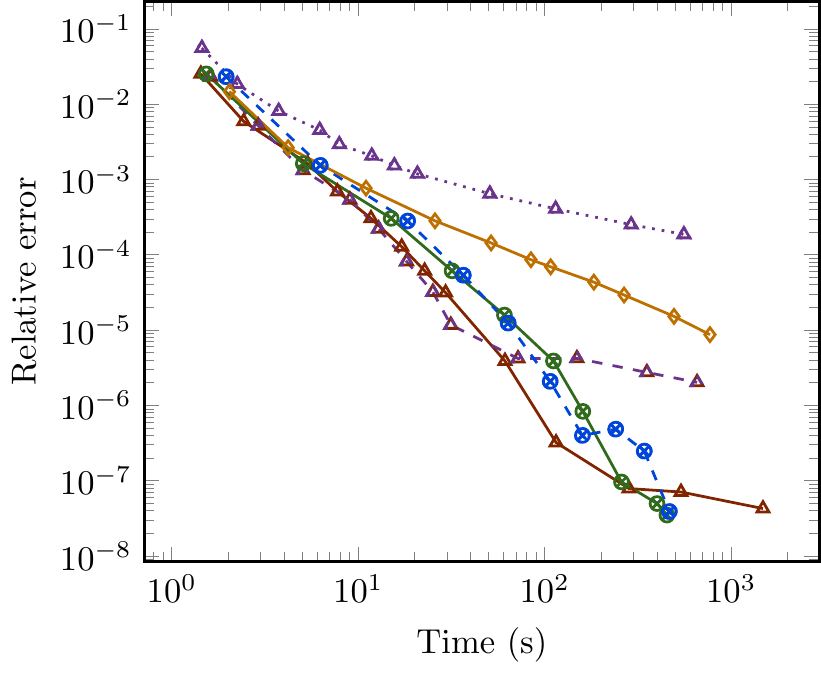}}
\caption{Comparison of the relative error of the 2-norm of the scalar flux across the domain, for different angular discretisations, for the Brunner problem. The \textcolor{foliagegreen}{$\otimes$} are regular P$_n$ adapts with threshold coefficient 1\xten{-5} and reduced tolerance solves, with the dashed \textcolor{matlabblue}{$\otimes$} regular FP$_n$ adapts with threshold coefficient 1\xten{-4}, spatially dependent $\Sigma_{\textrm{f}}$, with $\Sigma_{\textrm{f}}^{\textrm{1}}=1$ and reduced tolerance solves. The solid \textcolor{fireenginered}{$\bigtriangleup$} is uniform P$_n$, with the dashed \textcolor{gaylordpurple}{$\bigtriangleup$} FP$_n$ with $\Sigma_{\textrm{f}}=1$, the dotted \textcolor{gaylordpurple}{$\bigtriangleup$} FP$_n$ with $\Sigma_{\textrm{f}}=100$ and \textcolor{deludedorange}{$\diamond$} uniform LS P$^0$ FEM.}
\label{fig:brunner_result}
\end{figure}

\fref{fig:brunner_result} compares the best performing results from the adapted P$_n$ and adapted FP$_n$ with spatially dependent filter from Figures \ref{fig:brunner_adapt} and \ref{fig:brunner_adapt_fpn_space}, respectively, with a number of different discretisations taken from \cite{Dargaville2019}. We can see in \fref{fig:brunner_convg} that both the adapted P$_n$ and adapted FP$_n$ outperform all other discretisations per DOF. Indeed, in this problem the adapted FP$_n$ even outperforms the adapted P$_n$ method, which is an interesting result. This shows that even in a problem with sufficient smoothness that uniform P$_n$ exhibits good convergence when compared to a first order method like uniform LS P$^0$ FEM, there can still be a benefit to applying a filter. Both adapted P$_n$ and adapted FP$_n$ achieve similar runtime improvements over the uniform P$_n$ in this problem, as shown in \fref{fig:brunner_time}. This is the impact of the adaptivity reducing the number of large Riemann decompositions that must be performed. 
% ~~~~~~~~~~~~~
% ~~~~~~~~~~~~~

In particular, Table \ref{tab:brunner_memory} shows that although \fref{fig:brunner_result} suggests that the adapted FP$_n$ with spatially dependent filter is the best performing method in this problem, the cumulative runtime per final DOF is not constant, indicating that the overall method cannot be considered scalable (unlike the wavelet adaptivity in \cite{Dargaville2019} in this problem). This is entirely due to the increasing cost of performing/applying the Riemann decompositions, as Table \ref{tab:brunner_memory} shows that linear solver for our adapted FP$_n$ method actually requires fewer iterations as the maximum angular order is increased. Over half the overall runtime for the 10th adapt step is spent computing Riemann decompositions. The overall time spent computing these is still much less than if we were running a uniform simulation (as shown in \fref{fig:brunner_time}), but it is still the main contributing factor in this problem. Thankfully however, Table \ref{tab:brunner_memory} shows that the block form of $\mat{D}^{-1}$ means our peak memory use remains constant as the angular order is increased, settling down to approximately 37 copies of the angular flux (much like \cite{Dargaville2019} this can be reduced considerably by decreasing our GMRES restart parameter and computing $\mat{D}^{-1}$ on the fly with little impact on convergence/runtimes). We now turn to a problem with much larger discontinuities, where we would expect P$_n$ to perform poorly.
%~~~~~~~~~~~~~~~~
\begin{table}
\centering
\begin{tabular}{ l c c c c c c c c c c c}
\toprule
\textbf{Adapt step (max. order):} & \textbf{1 (1)} & \textbf{2 (5)} & \textbf{3 (9)} & \textbf{4 (13)} & \textbf{5 (17)} & \textbf{6 (21)} & \textbf{7 (25)} & \textbf{8 (29)} & \textbf{9 (33)} & \textbf{10 (37)}\\\midrule  
Cum. runtime ($\mu$s) per final DOFs: & 57.2 & 30.6 & 35.2 & 38.5 & 43.6 & 52.6 & 61.7 & 80.3 & 104.1 & 132.9 \\
No. iterations: & 12 & 14 & 20 & 20 & 19 & 19 & 19 & 18 & 18 & 17 \\
\% runtime Riemann decomp.: & 0.1 & 0.8 & 1.3 & 3 & 6.7 & 12.7 & 21.7 & 33.3 & 45.3 & 55.9 \\
Peak memory use: & 201.2 & 61.1 & 44.2 & 39.5 & 37.6 & 36.7 & 36.7 & 37 & 37.5 & 37.2\\
\bottomrule  
\end{tabular}
\caption{Runtime, iteration count, percentage of runtime spent computing Riemann decompositions and peak memory used for the Brunner problem, for the regular FP$_n$ adapt with threshold coefficient 1\xten{-4}, spatially dependent $\Sigma_{\textrm{f}}^{\textrm{1}}=1$ and reduced tolerance solves. Peak memory use is on the heap (measured by massif) scaled to the size of the angular flux. The runtime is the cumulative runtime of all adapt steps up to that level, scaled by the NDOFs in the final adapt step. The iteration count is from the linear solve at the final adapt step.}
\label{tab:brunner_memory}
\end{table}
%~~~~~~~~~~~~~~~~

% ~~~~~~~~~~~~~
\subsection{2D dogleg problem}
\label{sec:2D dogleg problem}
% ~~~~~~~~~~~~~
We now examine the use of spherical harmonics adaptivity in a 2D duct problem \cite{Goffin2015a}. We discretise this problem in space with the same mesh as \cite{Dargaville2019}, an unstructured triangular mesh with 2824 elements (1477 CG nodes and 8472 DG nodes). This problem features highly anisotropic flux throughout the duct regions and features heavy streaming region. This problem is highly challenging for high-order methods like P$_n$, with heavy discontinuities in space/angle. We use goal-based adaptivity in this problem, with the goal being the average flux at the end of the duct. The reference solution for the P$_n$ and FP$_n$ is uniform P$_{91}$ with 4278 DOFs in angle, using 42M DOFs. 

\begin{figure}[th]
\centering
\subfloat[][Error vs CDOFs]{\label{fig:dogleg_convg_fpn}\includegraphics[width =0.47\textwidth]{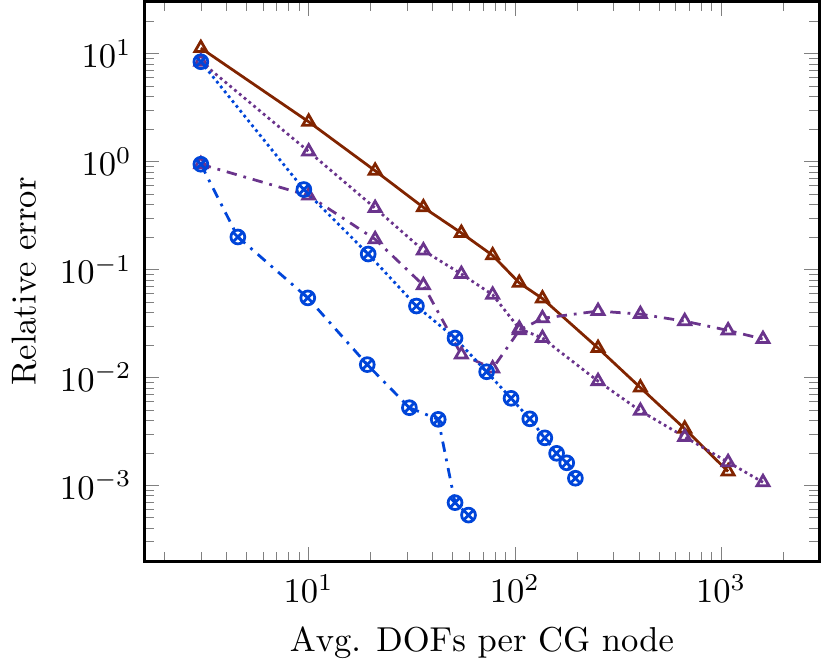}}
\subfloat[][Error vs total runtime]{\label{fig:doglethreshold coefficientg_time_fpn}\includegraphics[width =0.47\textwidth]{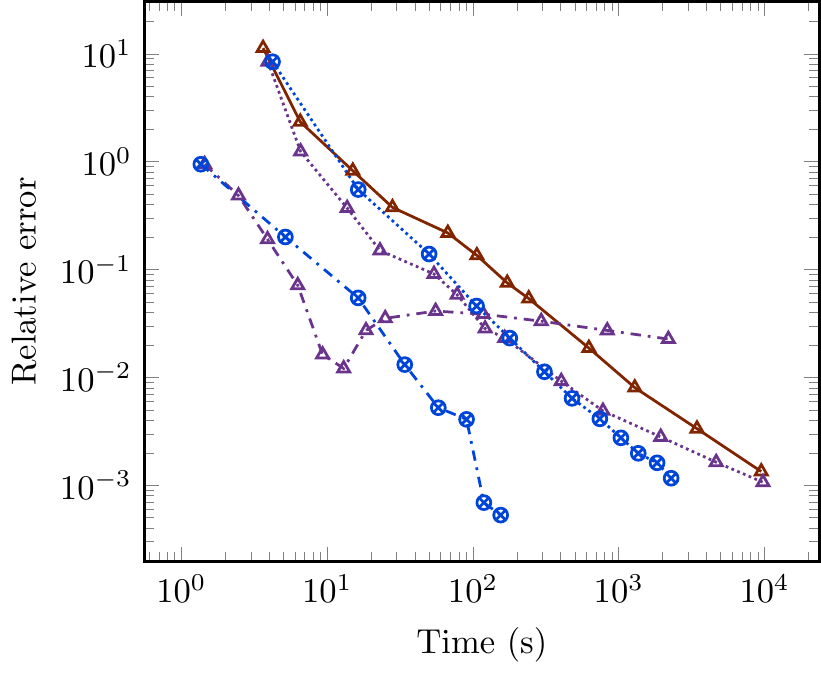}}\\
\caption{Effect of changing the filter strength of FP$_n$, in the relative error of the detector response, for the 2D dogleg problem. The solid \textcolor{fireenginered}{$\triangle$} is uniform P$_n$ and the \textcolor{gaylordpurple}{$\triangle$} are uniform FP$_n$, with densely dotted $\Sigma_{\textrm{f}}=0.1$ and dash dotted $\Sigma_{\textrm{f}}=10$. The \textcolor{matlabblue}{$\otimes$} are goal-based FP$_n$ adapts, error target 1\xten{-1} and reduced tolerance solves, spatially dependent $\Sigma_{\textrm{f}}$ with densely dotted $\Sigma_{\textrm{f}}^{\textrm{1}}=0.1$ and dash-dotted $\Sigma_{\textrm{f}}^{\textrm{1}}=10$}
\label{fig:dogleg_fpn}
\end{figure}

In this example we move directly to examining the impact of different filter values, the choice of the error tolerance used in this section (1\xten{-1}) was made following a similar process to that shown in \secref{sec:Brunner lattice problem} and \cite{Dargaville2019}. \fref{fig:dogleg_convg_fpn} shows that in this problem, uniform FP$_n$ with a constant filter value improves convergence per DOF for low order, but that there is still sufficient smoothness that at high order the uniform P$_n$ method converges better. We can see that overfiltering with $\Sigma_{\textrm{f}}=10$ also results in non-monotonic convergence in this  problem. One of the key features of \fref{fig:doglethreshold coefficientg_time_fpn} however is that the uniform FP$_n$ with constant filter values significantly outperform the P$_n$ method in runtime until high order. For example, achieving an error of approximately 1\xten{-2} with FP$_n$ with $\Sigma_\textrm{f}=10$ took only 13 seconds, with the $P_n$ taking around 1000 seconds. This is because the conditioning in this problem is improved considerably by filtering; Table \ref{tab:dogleg_result} shows that the iteration count for uniform P$_{45}$ is 968, compared with only 41 for uniform FP$_{45}$ with $\Sigma_\textrm{f}=10$. Table \ref{tab:dogleg_result} does show that the iteration count for FP$_n$ with $\Sigma_\textrm{f}=10$ is still growing slowly however as the angular order is increased. 
% ~~~~~~~~~~~~~
% ~~~~~~~~~~~~~
%~~~~~~~~~~~~~~~~~~~~~~~~~~~~~~~~~~~~
\begin{table}[ht]
\centering
\begin{tabular}{ l c c c c c c c c c c c c}
\toprule
\textbf{Order:} & \textbf{1} & \textbf{3} & \textbf{5} & \textbf{7} & \textbf{9} & \textbf{11} & \textbf{13} & \textbf{15} & \textbf{21} & \textbf{27} & \textbf{35} & \textbf{45}\\
\midrule  
Uniform P$_n$ & 47 & 62 & 95 & 119 & 207 & 244 & 277 & 309 & 420 & 511 & 741 & 968 \\
Uniform FP$_n$, $\Sigma_{\textrm{f}}=0.1$ & 38 & 45 & 63 & 73 & 127 & 141 & 156 & 171 & 225 & 269 & 353 & 482 \\
Uniform FP$_n$, $\Sigma_{\textrm{f}}=10$ & 9 & 13 & 14 & 16 & 20 & 21 & 22 & 24 & 27 & 31 & 36 & 41 \\
\midrule
\textbf{Adapt step (max. order):} & \textbf{1 (1)} & \textbf{2 (5)} & \textbf{3 (9)} & \textbf{4 (13)} & \textbf{5 (17)} & \textbf{6 (21)} & \textbf{7 (25)} & \textbf{8 (29)} & \textbf{9 (33)} & \textbf{10 (37)} & \textbf{11 (41)} & \textbf{12 (45)}\\
\midrule  
Goal-based P$_n$ & 47 & 89 & 165 & 190 & 223 & 274 & 308 & 351 & 400 & 421 & 458 & - \\
Goal-based FP$_n$, $\Sigma_{\textrm{f}}^{\textrm{1}}=10$ & 9 & 32 & 66 & 82 & 82 & 83 & 88 & 85 & - & - & - & -\\
\bottomrule  
\end{tabular}
\caption{Number of iterations for the different uniform/adapted discretisations shown in \fref{fig:dogleg_result}, for the 2D dogleg problem. For the adapted results, we take the iteration count from the last forward linear solve, which is always performed to the same tolerance as the uniform.}
\label{tab:dogleg_result}
\end{table}
% ~~~~~~~~~~~~~

\begin{figure}[th]
\centering
\subfloat[][Number of angular basis functions across the spatial domain (log scale) - 603K DOFs]{\label{fig:dogleg_fpn_no_angles_10_steps_1e-4_space_red}\includegraphics[width =0.47\textwidth]{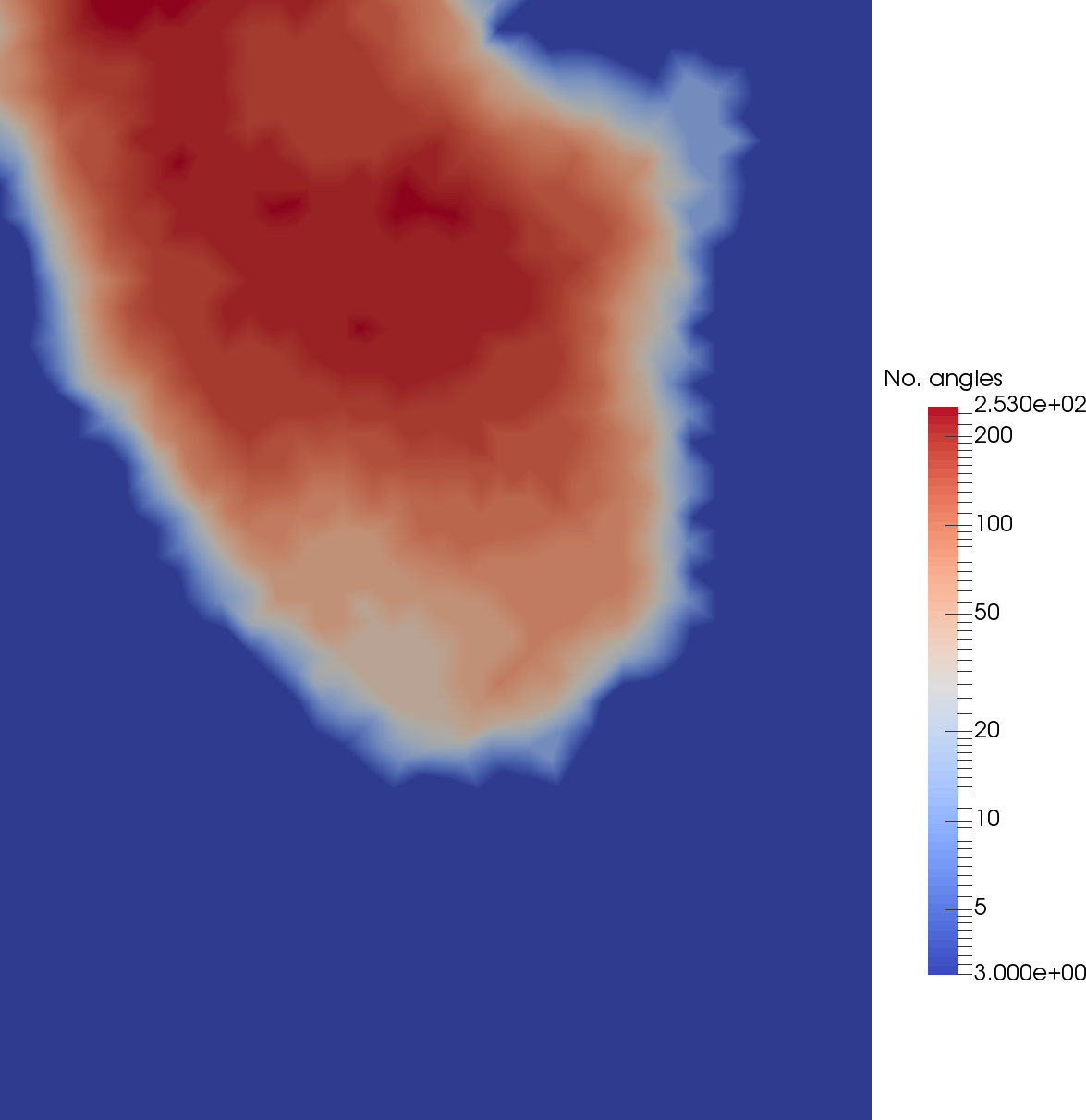}}
\subfloat[][Scalar flux across the domain]{\label{fig:dogleg_fpn_flux_10_steps_1e-4_space_red}\includegraphics[width =0.47\textwidth]{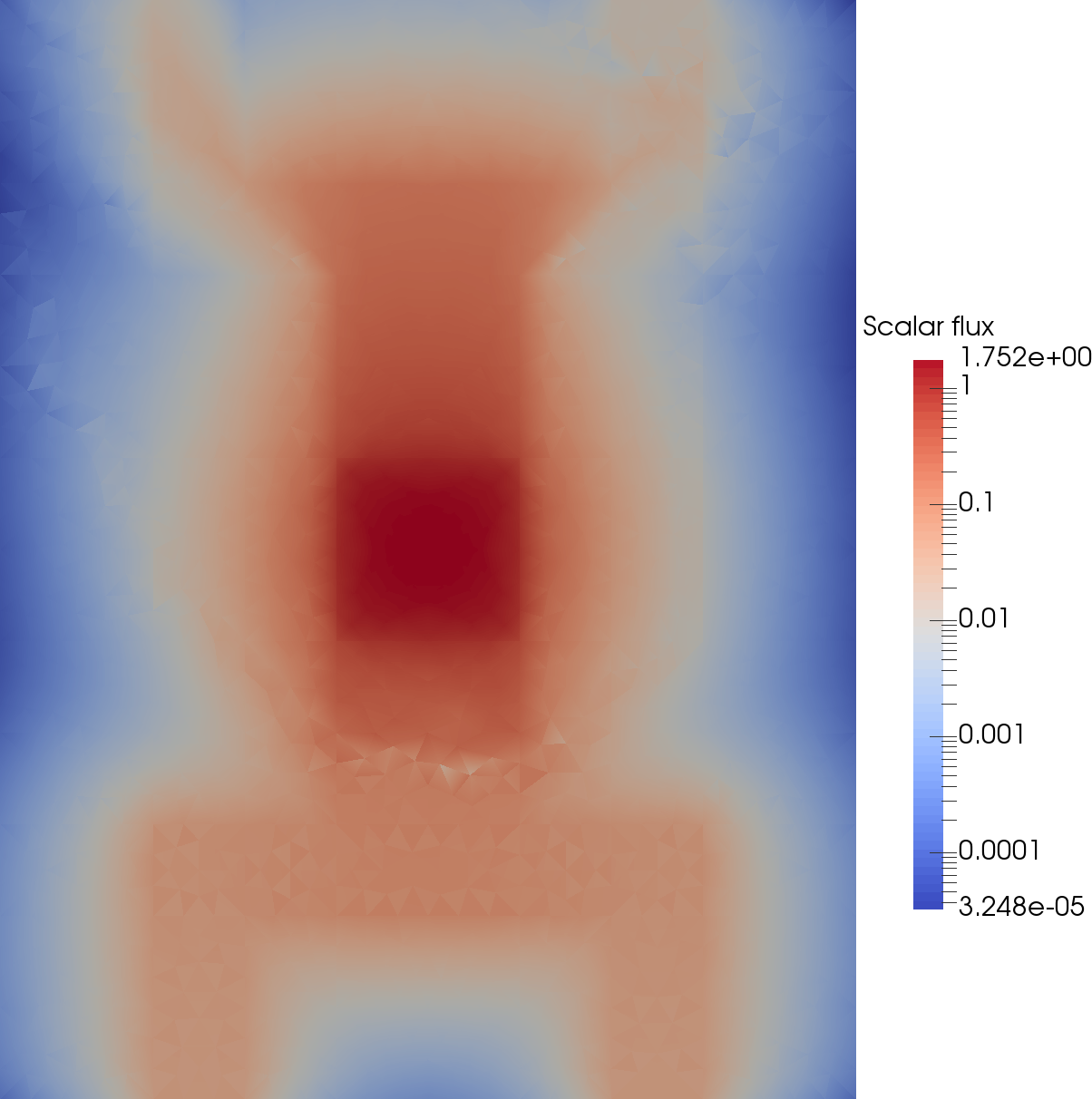}}\\
\subfloat[][Scalar flux in the adjoint]{\label{fig:dogleg_fpn_flux_adjoint_10_steps_1e-4_space_red}\includegraphics[width =0.47\textwidth]{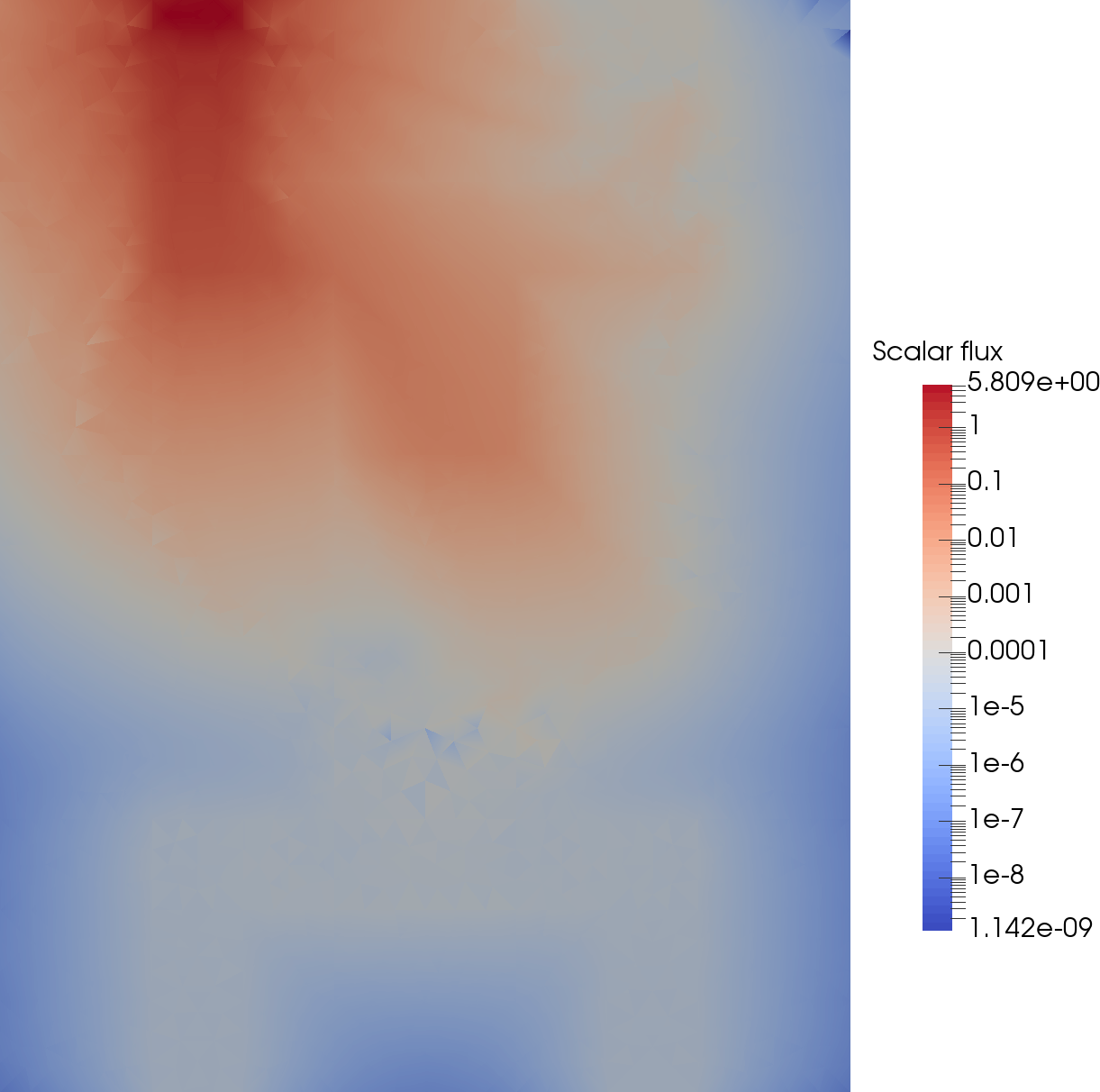}}
\caption{Results from the goal-based adaptivity with FP$_n$ on the 8th adapt step with error target 1\xten{-1} for the 2D dogleg problem, with spatially dependent $\Sigma_{\textrm{f}}$, with $\Sigma_{\textrm{f}}^{\textrm{1}}=10$.}
\label{fig:dogleg_flux}
\end{figure}

Importantly, \fref{fig:dogleg_fpn} also shows the results from allowing our FP$_n$ method to adapt and use a spatially dependent filter. We can see that this combination reduces both the average NDOFs per CG node and the runtime in this problem by at least two orders of magnitude. As in the previous example problem, allowing the filter to become spatially dependent improves the convergence when compared to the constant filter case. \fref{fig:dogleg_flux} shows the scalar flux for both the forward and adjoint problem, along with where the goal-based adaptivity has increased the angular order in this problem. \fref{fig:dogleg_fpn_no_angles_10_steps_1e-4_space_red} shows that the highest angular orders have been applied in the streaming path between the source and goal, as would be expected. One important feature to note in the forward/adjoint scalar fluxes shown in Figures \ref{fig:dogleg_fpn_flux_10_steps_1e-4_space_red} and \ref{fig:dogleg_fpn_flux_adjoint_10_steps_1e-4_space_red} is that the solution is entirely positive, and as mentioned in \secref{sec:Spatially dependent filter}, we can see discontinuities where the angular resolution has transitioned from P$_1$. 
% ~~~~~~~~~~~~~
% ~~~~~~~~~~~~~

\begin{figure}[th]
\centering
\subfloat[][Absolute value of $\bm{\Sigma}_{\textrm{stab}}$ across space. This can be considered the net ``amount'' of stabilisation we apply at each node.]{\label{fig:dogleg_fpn_stab_10_steps_1e-4_space_red}\includegraphics[width =0.47\textwidth]{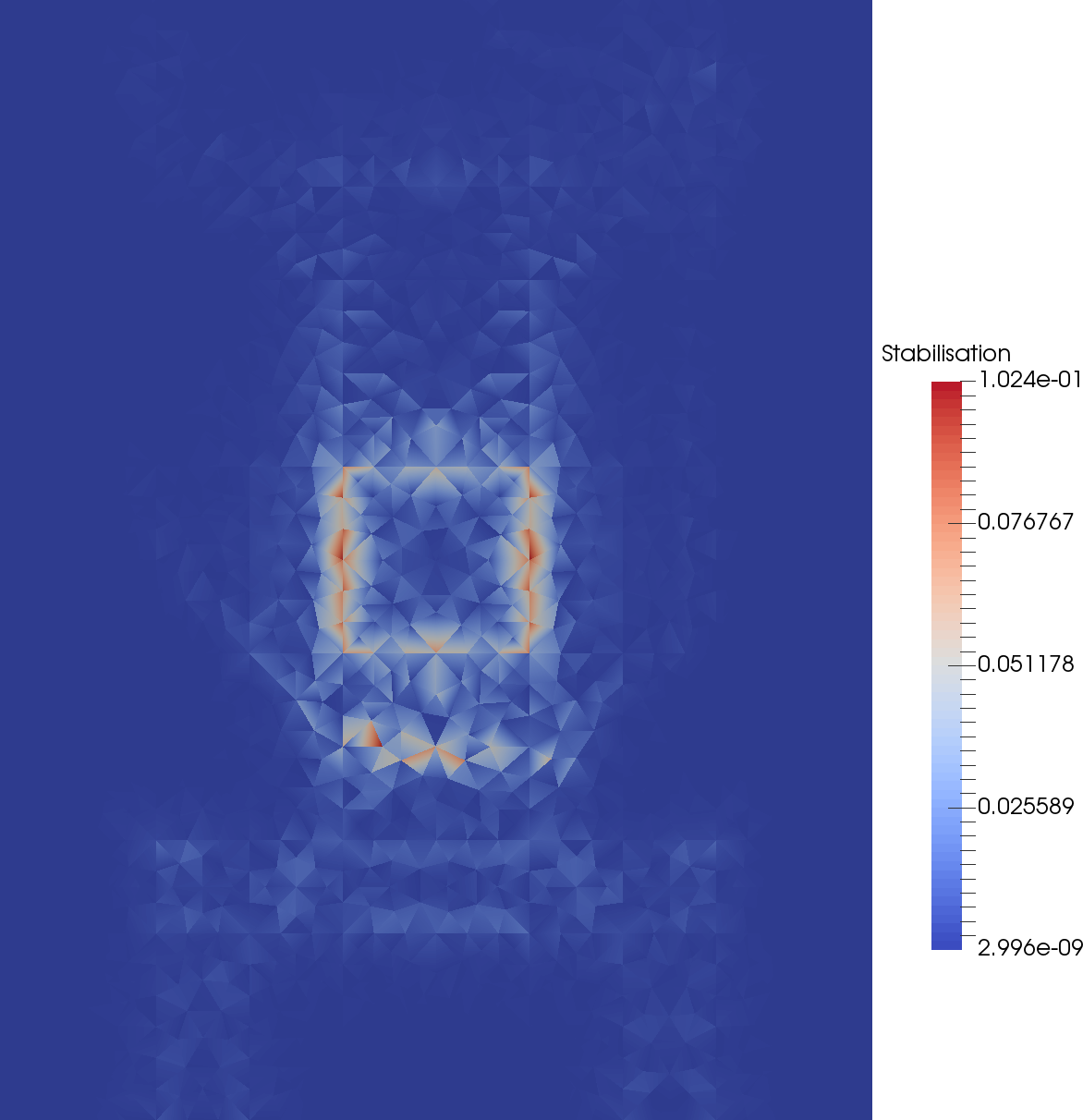}}\hspace{0.1cm}
\subfloat[][Spatially-dependent $\Sigma_{\textrm{f}}$]{\label{fig:dogleg_fpn_filter_10_steps_1e-4_space_red}\includegraphics[width =0.47\textwidth]{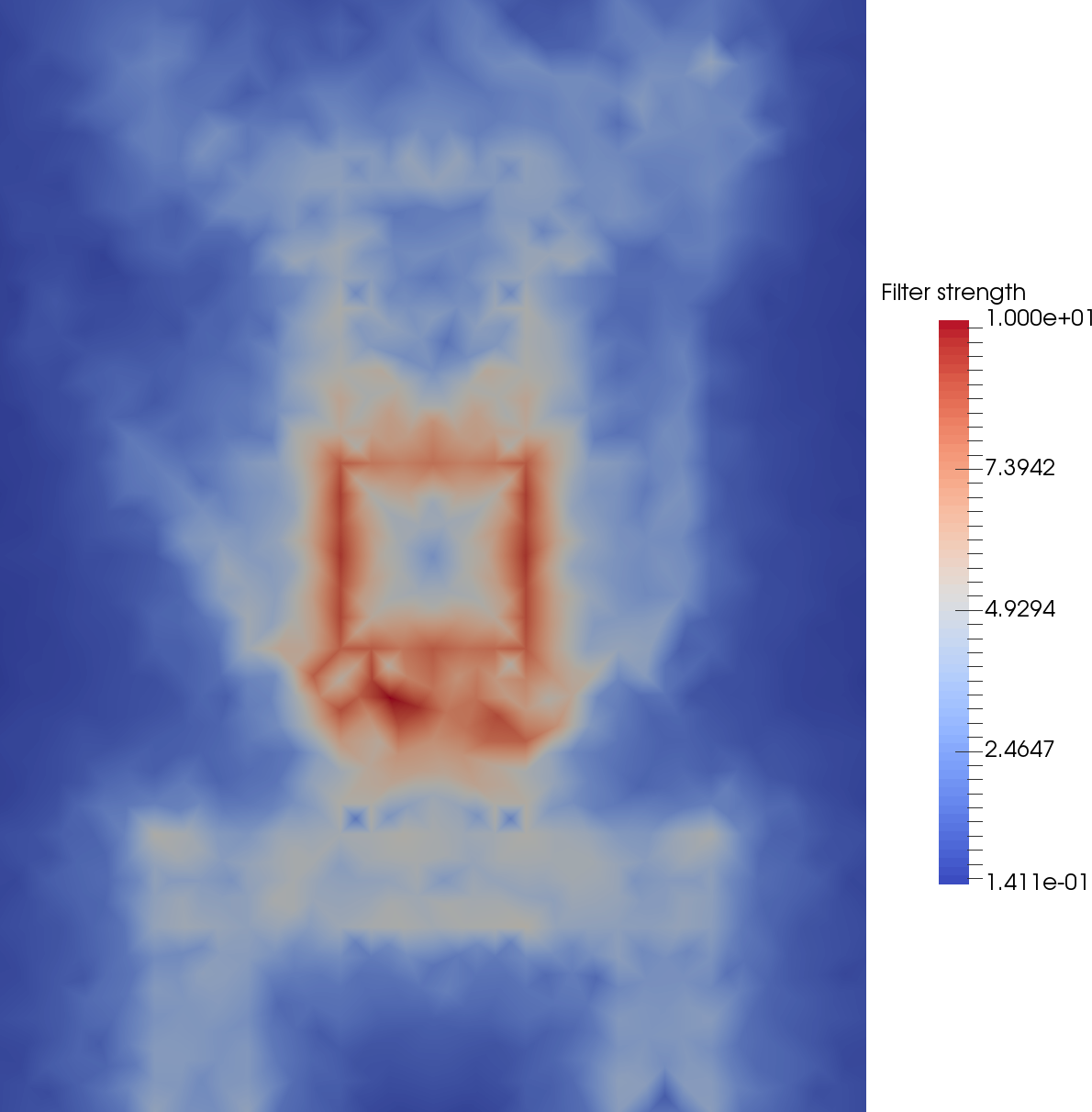}}\\
\subfloat[][Absolute value of $\bm{\Sigma}_{\textrm{stab}}$ across space for the adjoint (log scale). This can be considered the net ``amount'' of stabilisation we apply at each node.]{\label{fig:dogleg_fpn_stab_adjoint_10_steps_1e-4_space_red}\includegraphics[width =0.47\textwidth]{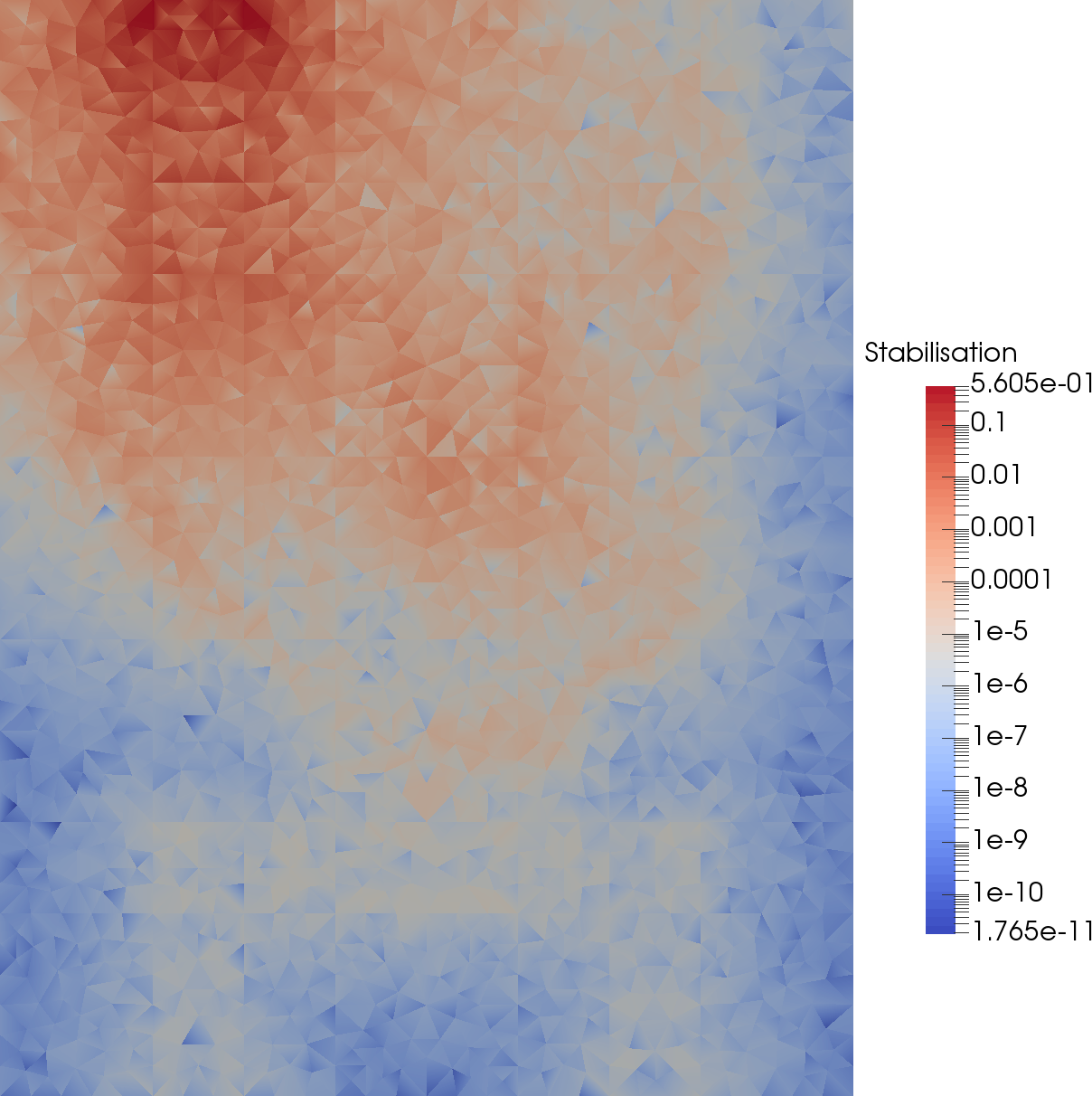}}\hspace{0.1cm}
\subfloat[][Spatially-dependent $\Sigma_{\textrm{f}}$ in the adjoint (log scale)]{\label{fig:dogleg_fpn_filter_adjoint_10_steps_1e-4_space_red}\includegraphics[width =0.47\textwidth]{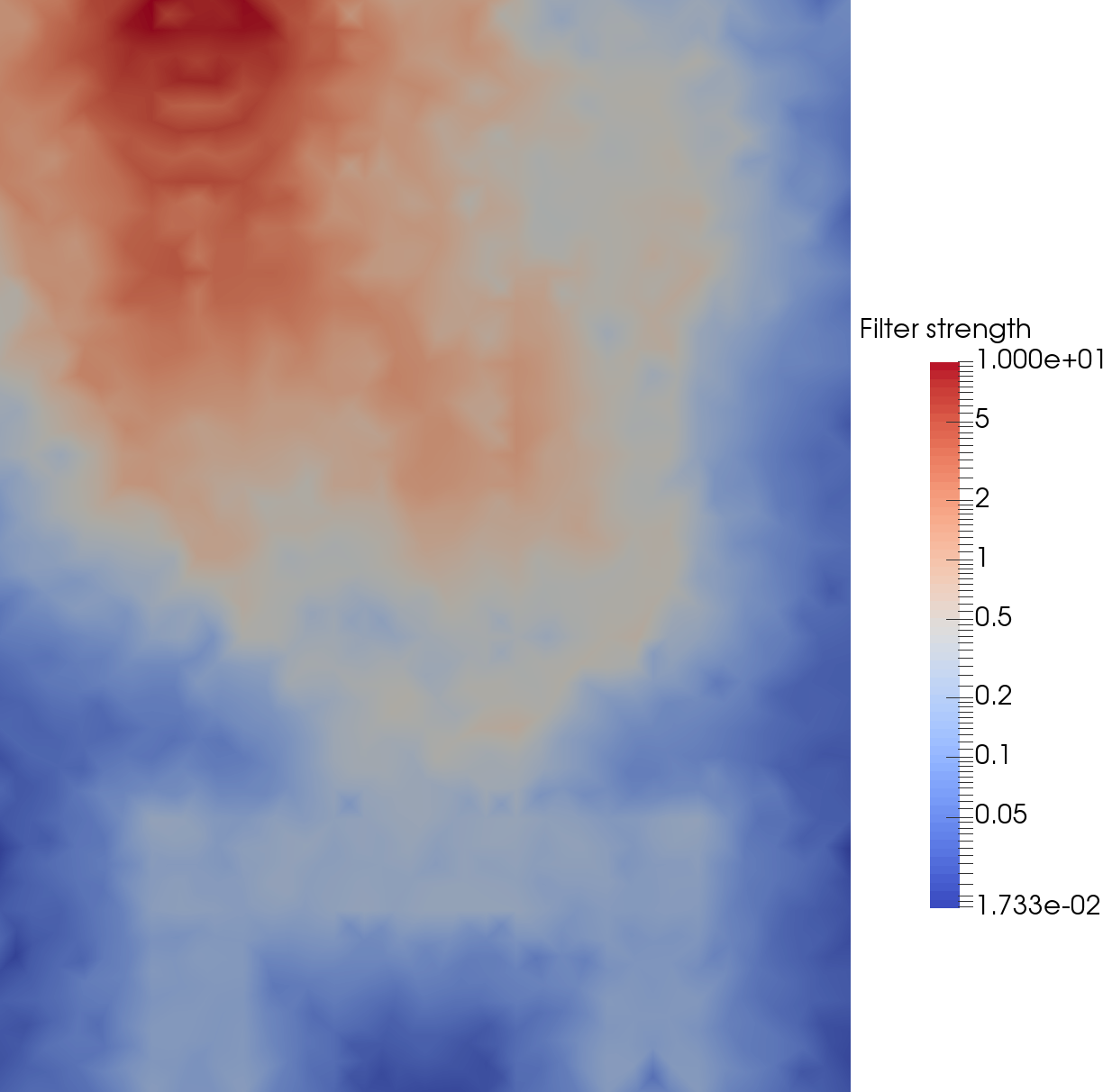}}\\
\caption{Computation of a spatially dependent $\Sigma_{\textrm{f}}$, with $\Sigma_{\textrm{f}}^{\textrm{1}}=10$, on the the 8th step of FP$_n$ goal-based adaptivity with error target 1\xten{-1} for the 2D dogleg problem.}
\label{fig:dogleg_space}
\end{figure}

Thankfully, as discussed this discontinuity is picked up by $\Sigma_\textrm{stab}$ shown in Figures \ref{fig:dogleg_fpn_stab_10_steps_1e-4_space_red} and \ref{fig:dogleg_fpn_stab_adjoint_10_steps_1e-4_space_red} in the forward and adjoint solutions respectively. This means our spatially dependent filter is also filtering heavier where this change in angular resolution has caused discontinuities, as shown in Figures \ref{fig:dogleg_fpn_filter_10_steps_1e-4_space_red} and \ref{fig:dogleg_fpn_filter_adjoint_10_steps_1e-4_space_red}. The other large discontinuities in this problem are present around the sources and in the duct regions, as expected. 
% ~~~~~~~~~~~~~
% ~~~~~~~~~~~~~

\begin{figure}[th]
\centering
\subfloat[][Error vs CDOFs]{\label{fig:dogleg_convg}\includegraphics[width =0.47\textwidth]{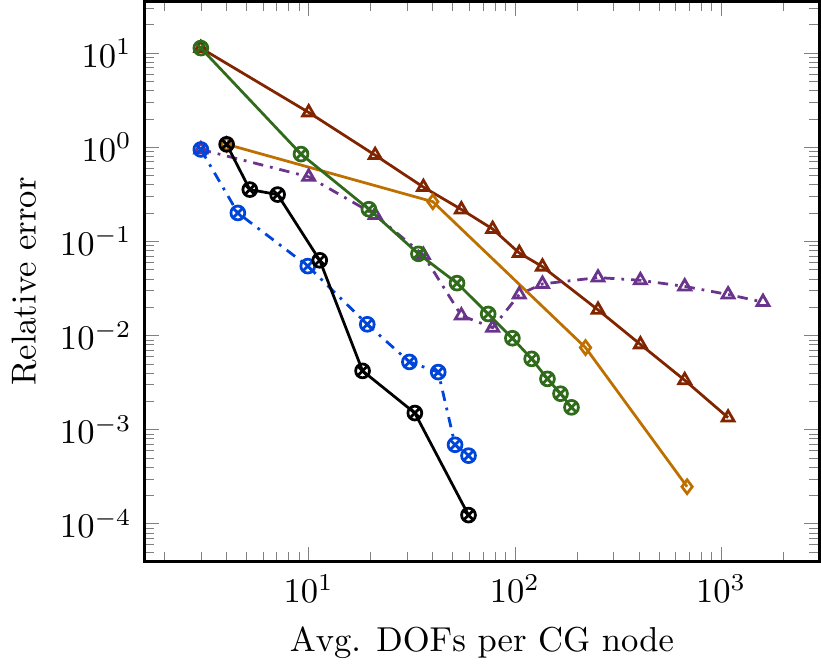}}
\subfloat[][Error vs total runtime]{\label{fig:dogleg_time}\includegraphics[width =0.47\textwidth]{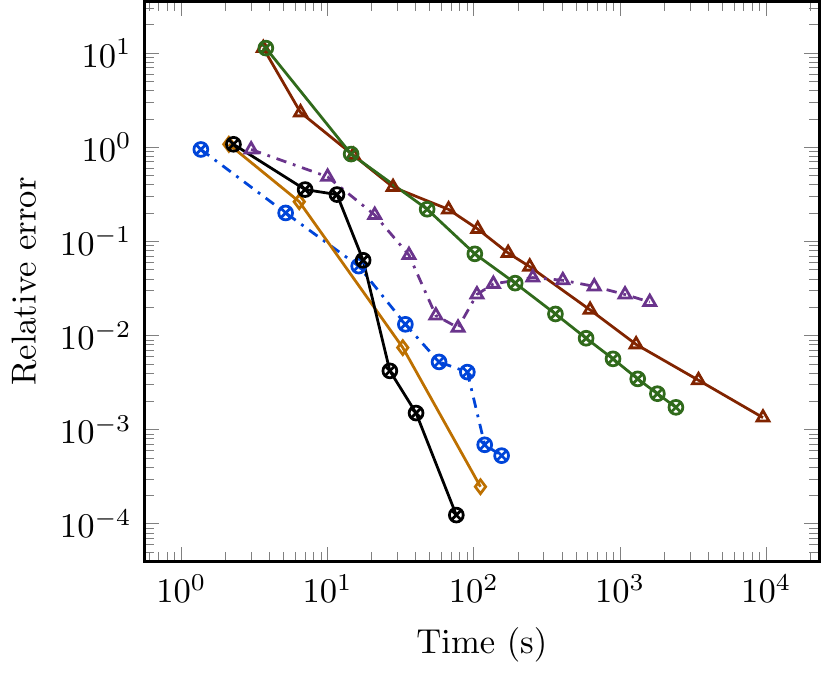}}
\caption{Comparison of the relative error of the 2-norm of the scalar flux across the domain, for different angular discretisations, for the 2D dogleg problem. The \textcolor{foliagegreen}{$\otimes$} are goal-based P$_n$ adapts with error target 1\xten{-1} and reduced tolerance solves, with the dashed \textcolor{matlabblue}{$\otimes$} goal-based FP$_n$ adapts with error target 1\xten{-1}, spatially dependent $\Sigma_{\textrm{f}}$, with $\Sigma_{\textrm{f}}^{\textrm{1}}=10$ and reduced tolerance solves. The solid \textcolor{fireenginered}{$\bigtriangleup$} is uniform P$_n$, the dash-dotted \textcolor{gaylordpurple}{$\triangle$} is uniform FP$_n$ with $\Sigma_{\textrm{f}}=10$ and \textcolor{deludedorange}{$\diamond$} uniform LS P$^0$ FEM. The \textcolor{black}{$\otimes$} are goal-based adapted non-standard Haar wavelets with error target 1\xten{-3} and one extra adapt step (from \cite{Dargaville2019})}
\label{fig:dogleg_result}
\end{figure}

We now compare the results from both the goal-based adapted P$_n$ and filtered P$_n$, along with a range of different angular discretisations from \cite{Dargaville2019}. We can see in \fref{fig:dogleg_convg} adapated P$_n$ in this problem reduces NDOFs by an order of magnitude in this problem when compared to the uniform P$_n$, and results in a drop in runtime at high order, shown in \fref{fig:dogleg_time}. Table \ref{tab:dogleg_result} shows that the the adapted P$_n$ has resulted in a smaller number of iterations in the linear solver when compared to uniform P$_n$. The adapted FP$_n$ with spatially dependent filter gives excellent results in this problem, giving a further order of magnitude reduction in DOFs, and due to the improved conditioning, almost two orders of magnitude decrease in runtime when compared to the uniform P$_n$. Again we can see this reflected in the iteration count in Table \ref{tab:dogleg_result}, and importantly we see a constant iteration count with increasing maximum angular order. This is a powerful result for a filtered spectral method in a streaming problem. Indeed \fref{fig:dogleg_result} shows that the goal-based adaptive FP$_n$ with spatially dependent filter is even competitive with the scalable goal-based P$^0$ wavelet calculation from \cite{Dargaville2019}, both in NDOFs and runtime.
% ~~~~~~~~~~~~~
% ~~~~~~~~~~~~~

This behaviour will not continue as the maximum adaptive order of our FP$_n$ method increases given the non-scalable computation/application of Riemann conditions discussed above, whereas the wavelets remain scalable given any level of refinement. Indeed we see this in Table \ref{tab:dogleg_memory}, where the cumulative runtime per DOF is starting to grow slowly with angular order for the adaptive FP$_n$. Given we have a fixed iteration count, we know this is due to the Riemann decompositions, and we can see in Table \ref{tab:dogleg_memory} the overall percentage runtime spent computing these decompositions also rising. Thankfully again we see the peak memory consumption for our adaptive method is scalable however, stabilising at around 63 copies of the angular flux (which is slightly less than twice that shown in Table \ref{tab:brunner_memory}, given we solve a forward and adjoint problem and benefit from reuse of some data structures).
%~~~~~~~~~~~~~~~~
\begin{table}
\centering
\begin{tabular}{ l c c c c c c c c c}
\toprule
\textbf{Adapt step (max. order):} & \textbf{1 (1)} & \textbf{2 (5)} & \textbf{3 (9)} & \textbf{4 (13)} & \textbf{5 (17)} & \textbf{6 (21)} & \textbf{7 (25)} & \textbf{8 (29)}\\\midrule  
Cum. runtime ($\mu$s) per final DOFs: & 45.3 & 113.5 & 162.6 & 174 & 184.6 & 209.2 & 228.2 & 256.6\\
\% runtime Riemann decomp.: & 0.002 & 0.8 & 1.4 & 1.8 & 3.3 & 4.8 & 5.5 & 6.5 \\
Peak memory use: & 266.6 & 239.3 & 127.9 & 86.8 & 71.6 & 64.8 & 63.2 & 63.5\\
\bottomrule  
\end{tabular}
\caption{Runtime, percentage of runtime spent computing Riemann decompositions and peak memory used for the 2D dogleg problem, for the goal-based FP$_n$ adapt with error target 1\xten{-1}, spatially dependent $\Sigma_{\textrm{f}}^{\textrm{1}}=10$ and reduced tolerance solves. Peak memory use is on the heap (measured by massif) scaled to the size of the angular flux. The runtime is the cumulative runtime of all adapt steps up to that level, scaled by the NDOFs in the final adapt step.}
\label{tab:dogleg_memory}
\end{table}
%~~~~~~~~~~~~~~~~

Finally in this problem, we examine the effectivity index of our goal-based error metrics. In \cite{Dargaville2019}, we showed evidence of a scalable error metric that despite producing excellent results, had a pathological effectivity index. The improved error metric discussed in this work removes this pathology; Table \ref{tab:dogleg_effec} shows that for our goal-based adaptive P$_n$ method our effectivity index varies between 0.04 to 4.52 throughout the adapt process. This is due to both the improved error metric, but also the the absence of ray-effects in our spherical harmonics solution. This is an important result and we will examine this further in future work. Table \ref{tab:dogleg_effec} also shows that the effectivity index for our adapted FP$_n$ with spatially dependent filter is much worse, increasing to around 22 in the 7th adapt step. This is also not pathological and the results from \fref{fig:dogleg_result} indicate that even with a worse effectivity index than the adapted P$_n$, our error metric still results in refinement in correct regions, much like in \cite{Dargaville2019}. Given the discussion in \secref{sec:Goal-based adaptivity}, part of this poorer effectivity index is due to ignoring the differing filter values across space in the forward and adjoint problems, and if we wish to improve our effectivity index in this case we would simply need to project into a common space; we leave examining this to future work. 
%~~~~~~~~~~~~~~~~~~~~~~~~~~~~~~~~~~~~
\begin{table}[ht]
\centering
\begin{tabular}{ l c c c c c c c c c c c c}
\toprule
\textbf{Adapt step (max. order):} & \textbf{1 (1)} & \textbf{2 (5)} & \textbf{3 (9)} & \textbf{4 (13)} & \textbf{5 (17)} & \textbf{6 (21)} & \textbf{7 (25)} & \textbf{8 (29)} & \textbf{9 (33)} & \textbf{10 (37)} & \textbf{11 (41)}\\
\midrule  
Goal-based P$_n$ & 0.48 & 4.52 & 3.76 & 1.58 & 1.06 & 1.32 & 0.08 & 0.04 & 1.33 & 0.74 & 1.52 \\
Goal-based FP$_n$, $\Sigma_{\textrm{f}}^{\textrm{1}}=10$ & 0.48 & 8.75 & 16.4 & 10.1 & 13.3 & 6.17 & 22.46 & 11.79 & - & - & -\\
\bottomrule  
\end{tabular}
\caption{Effectivity index for the goal-based adapted discretisations shown in \fref{fig:dogleg_result}, for the 2D dogleg problem.}
\label{tab:dogleg_effec}
\end{table}
% ~~~~~~~~~~~~~
% ~~~~~~~~~~~~~
\subsection{3D void problem}
\label{sec:3D void problem}
% ~~~~~~~~~~~~~
% ~~~~~~~~~~~~~
Our final example problem is a source/detector problem in 3D, with a pure vacuum in the duct. This is the same as that shown in \cite{Dargaville2019}, except we have shortened the length of the duct from 320cm to 100cm. We discretise this problem with an unstructured tetrahedral mesh with 31,542 elements (9,525 CG nodes and 126,168 DG nodes) and  (this is a different mesh from \cite{Dargaville2019} as we do not have ray-effects that we would like to keep aligned with the mesh/geometry). This is a challenging problem for a spherical harmonics method, given the pure streaming, while also being trivial given we know the intensity of radiation must fall off from a (point) source like $r^{-2}$. Unsurprisingly, we could not compute a numerical solution with our iterative method for P$_n$ in this problem, given the poor conditioning. As such, this is an excellent problem for understanding the behaviour of the FP$_n$ method in the streaming limit. We also produce a reference solution in this problem with the non-standard Haar wavelets from \cite{Dargaville2019}, with fixed refinement up to 14 levels between $\mu \in [-0.0099, 0.0099]$ and $\omega ∈ [1.561, 1.5807]$.  
% ~~~~~~~~~~~~~
\begin{figure}[th]
\centering
\subfloat[][$\Sigma_{\textrm{f}}=1$]{\label{fig:tube_3D_flux_filter_1}\includegraphics[width =0.47\textwidth]{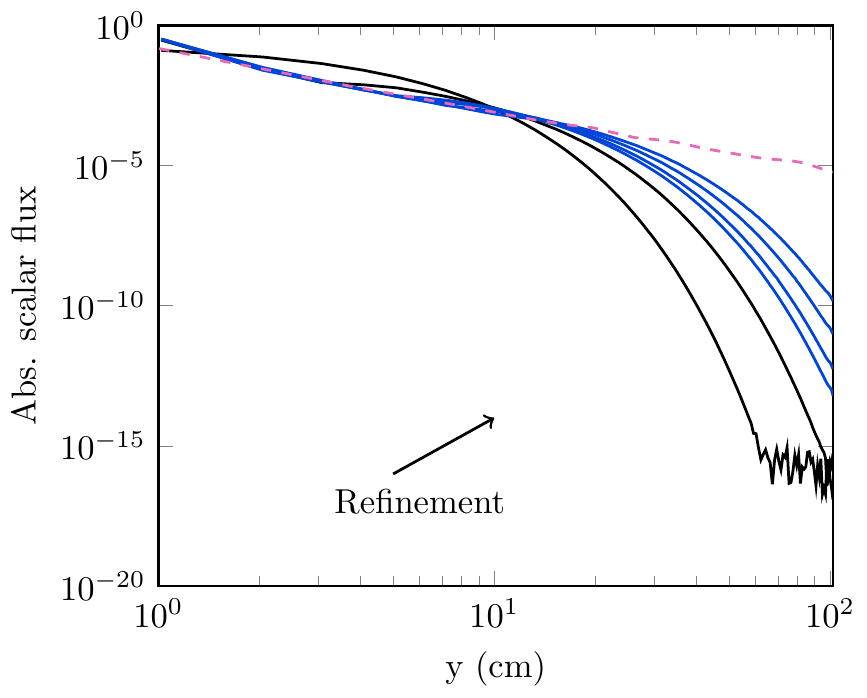}}
\subfloat[][$\Sigma_{\textrm{f}}=10$]{\label{fig:tube_3D_flux_filter_10}\includegraphics[width =0.47\textwidth]{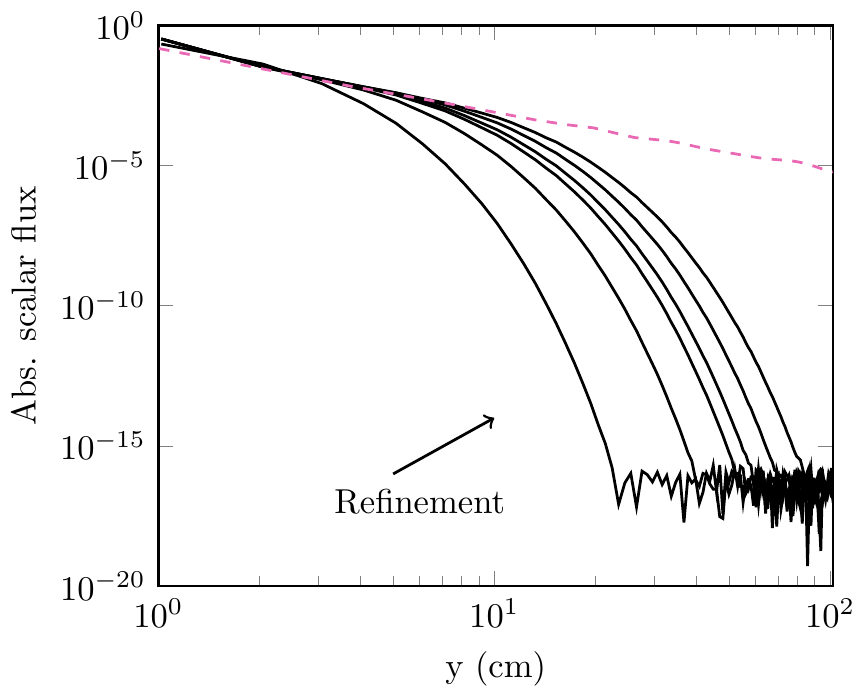}}
\caption{Absolute value of the scalar flux in the 3D duct along the length of the duct, from (0.5, 0.0, 0.5) to (0.5, 102, 0,5). The solid lines are FP$_n$, orders 1, 5, 9, 11, 15 and 21. The FP$_n$ solid lines are shaded blue when they are entirely positive down the length of the tube. The dashed pink line corresponds to the non-standard Haar wavelet discretisation in \cite{Dargaville2019} with 14 levels of refinement.}
\label{fig:tube_3D_filter}
\end{figure}
% ~~~~~~~~~~~~~

\fref{fig:tube_3D_filter} shows the results from uniform FP$_n$ simulations in this problem, with two different constant filter strengths. We can see in \fref{fig:tube_3D_flux_filter_1} that with constant filter of $\Sigma_{\textrm{f}}=1$, the FP$_9$ solution is the first to be both non-zero and positive along the entire length of the duct, with further refinement causing the solution to approach a highly refined P$^0$ solution (that agrees with the r$^{-2}$ drop-off). If we apply a stronger constant filter of $\Sigma_{\textrm{f}}=10$, we can see in \fref{fig:tube_3D_flux_filter_10} that even refinement up to FP$_{21}$ has failed to produce a non-zero solution at the end of the duct. This is to be expected, as heavier filtering smooths out peaks in the angular flux (by acting like ``negative'' forward-peaked scatter), making our solution less anisotropic. Of course this heavier filtering results in improved conditioning of our linear system and hence less iterations; Table \ref{tab:tube_3D_its} shows that the iteration count for FP$_{21}$ with $\Sigma_{\textrm{f}}=10$ is almost half that of $\Sigma_{\textrm{f}}=1$. Importantly, we can see that even with a filter of $\Sigma_{\textrm{f}}=1$, we have an almost constant iteration count with angular refinement in a problem with pure streaming. Although \fref{fig:tube_3D_flux_filter_1} shows that FP$_{21}$ with $\Sigma_{\textrm{f}}=1$ would not be considered high enough angular resolution to resolve this problem, it is perhaps surprising that a filtered spherical harmonics method can produce a non-zero solution (that is naturally free of ray-effects) in a duct with aspect ratio of 1:100 with such low resolution.
%~~~~~~~~~~~~~~~~~~~~~~~~~~~~~~~~~~~~
\begin{table}[ht]
\centering
\begin{tabular}{ l c c c c c c c c c c c c}
\toprule
\textbf{Order:} & \textbf{1} & \textbf{5} & \textbf{9} & \textbf{11} & \textbf{15} & \textbf{21}\\
\midrule  
Uniform P$_n$ & - & - & - & - & - & - \\
Uniform FP$_n$, $\Sigma_{\textrm{f}}=1$ & 46 & 61 & 179 & 185 & 196 & 207 \\
Uniform FP$_n$, $\Sigma_{\textrm{f}}=10$ & 32 & 35 & 92 & 95 & 102 & 112 \\
\midrule
\textbf{Adapt step (max. order):} & \textbf{-} & \textbf{-} & \textbf{1 (9)} & \textbf{2 (11)} & \textbf{3 (15)} & \textbf{4 (21)}\\
\midrule
Goal-based FP$_n$, $\Sigma_\textrm{f}^1=1$ & - & - & 179 & 231 & 244 & 250 \\
\bottomrule  
\end{tabular}
\caption{Number of iterations for the 3D void problem, solved to a relative tolerance of 1\xten{-14}. The P$_n$ simulations all diverged given the pure vacuum.}
\label{tab:tube_3D_its}
\end{table}
% ~~~~~~~~~~~~~

\begin{figure}[th]
\centering
\subfloat[][Scalar flux along the length of the duct, from (0.5, 0.0, 0.5) to (0.5, 102, 0,5). The blue lines are the forward problems, the green line is the adjoint problem on the 4th adapt step. The dashed pink line corresponds to the non-standard Haar wavelet discretisation in \cite{Dargaville2019} with 14 levels of refinement.]{\label{fig:tube_3D_flux_filter_1_adapt}\includegraphics[width =0.47\textwidth]{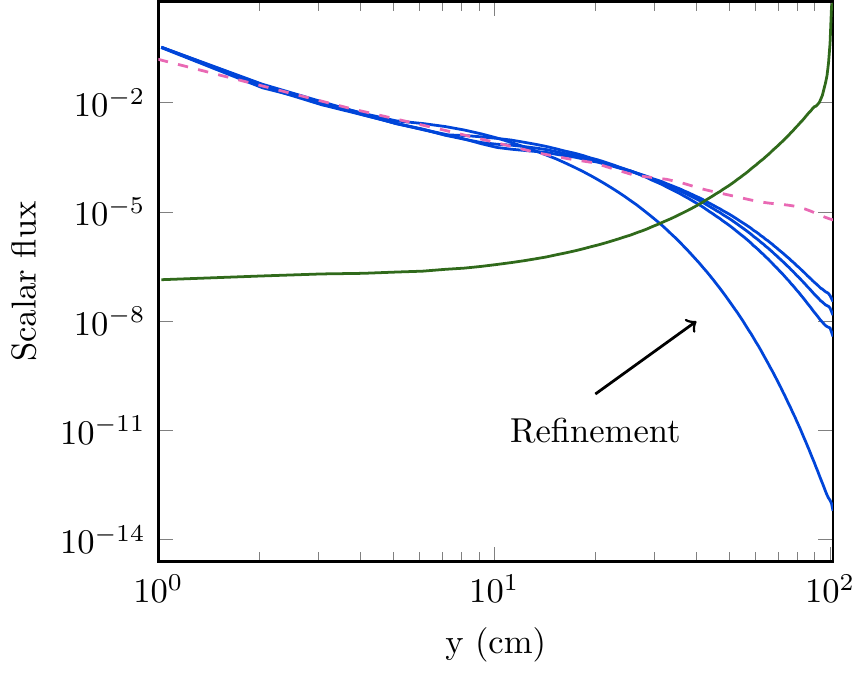}}\hspace{0.1cm}
\subfloat[][Filter strength along the length of the duct, from (0.5, 0.0, 0.5) to (0.5, 102, 0,5) on the 4th adapt step. The blue line is the forward problem, the green line is the adjoint.]{\label{fig:tube_3D_flux_filter_strength_adapt}\includegraphics[width =0.47\textwidth]{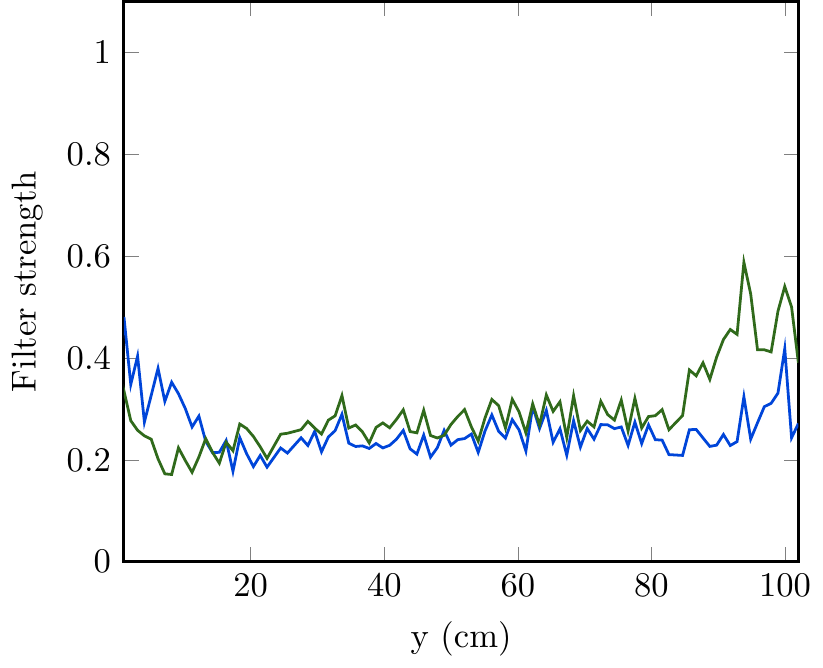}}
\caption{Results from goal-based FP$_n$ in the 3D duct problem with $\Sigma_{\textrm{f}}^{\textrm{1}}=1$, starting from FP$_9$ with four adapt steps.}
\label{fig:tube_3D_filter_adapt}
\end{figure}

If we are to apply our goal-based adaptivity in this problem, we must have a non-zero response in our functional at the coarsest angular resolution, as discussed in \secref{sec:Goal-based adaptivity}. These results indicate that for this 3D duct problem our FP$_9$ discretisation with $\Sigma_{\textrm{f}}^1=1$ would serve as a suitable coarse discretisation. We therefore use this as our coarse angular discretisation with our goal-based adaptivity and a spatially dependent filter value. In order to ensure our spatially dependent filter does not go to zero down the length of the duct (given the 15 orders of magnitude drop in scalar flux), as mentioned in \secref{sec:Spatially dependent filter}, we scale $\Sigma_\textrm{stab}$ for both the forward and adjoint problems by the respective scalar flux. \fref{fig:tube_3D_flux_filter_1_adapt} shows that our goal-based adaptivity successfully refines from our coarse FP$_9$ discretisation, producing a solution that is converging towards the refined P$^0$ solution (we also plot the refined adjoint solution to show it is similarly non-zero and positive). We don't plot the angular order at each spatial node in this problem, as the the goal-based adaptivity has resulted in angular refinement at every spatial node up to the max. order at each step, giving the same resolution as the uniform, as we might expect from a goal-based error metric in this problem.

% ~~~~~~~~~~~~~
% ~~~~~~~~~~~~~
\fref{fig:tube_3D_flux_filter_strength_adapt} shows the size of the spatially dependent filter down the length of the duct for both the forward and adjoint solutions, and we can see that the filter strength is almost constant. This implies that our net stabilisation drops off in a similar manner to the magnitude of the scalar flux, and that the scaled version of $\Sigma_\textrm{stab}$ is suitable for computing our $\Sigma_\textrm{f}$ in streaming problems. The magnitude has been reduced slightly from $\Sigma_\textrm{f}^1=1$ and this explains the improved solution shown from the FP$_{21}$ solution in \fref{fig:tube_3D_flux_filter_1_adapt} compared with that in \fref{fig:tube_3D_flux_filter_1}. This slight drop in filter strength does not heavily affect the iteration count with refinement, as shown in Table \ref{tab:tube_3D_its}, with the goal-based FP$_n$ method using 250 iterations in the final adapt step. Much like the unadapted case with constant filter strength, the iteration count is close to constant with angular refinement. We should also note that the cumulative runtime per final DOF does increase with adapt step in this problem, increasing from 596 $\mu$s, 1253 $\mu$s, 1917 $\mu$s to 2769 $\mu$s in the final adapt step. Note at such low order this is not due to computing the Riemann decompositions, which take only 1.5\% of the total runtime in the final adapt step (given our grouping, etc), but instead to applying them on each face. 

Table \ref{tab:tube_3D_effec} also shows the effectivity index for our goal-based FP$_n$ simulation, we can see that the our index is quite poor for early adapt steps, but improves considerably as we refine. This is to be expected in this problem, as Figures \ref{fig:tube_3D_filter} and \ref{fig:tube_3D_flux_filter_1_adapt} show our coarse angular solution is also poor near the end of the duct. The key point is that our effectivity index is non-zero, and \fref{fig:tube_3D_filter_adapt} shows it is causing refinement in the correct spatial regions. 
%~~~~~~~~~~~~~~~~~~~~~~~~~~~~~~~~~~~~
\begin{table}[ht]
\centering
\begin{tabular}{ l c c c c}
\toprule
\textbf{Adapt step (max. order):} & \textbf{1 (9)} & \textbf{2 (11)} & \textbf{3 (15)} & \textbf{4 (21)}\\
\midrule
Goal-based FP$_n$, $\Sigma_\textrm{f}^1=1$ & 8.5\xten{-9} & 6.3\xten{-3}& 2.0\xten{-2} & -4.7\xten{-2} \\
\bottomrule  
\end{tabular}
\caption{Effectivity index for the goal-based adapted discretisations shown in \fref{fig:dogleg_result}, for the 3D duct problem.}
\label{tab:tube_3D_effec}
\end{table}
% ~~~~~~~~~~~~~
% ~~~~~~~~~~~~~
\section{Conclusions}
% ~~~~~~~~~~~~~
This paper has presented an angular adaptivity algorithm for spherical harmonics that features both regular and goal-based error metrics, in problems with a range of smoothness. In particular, we examined the use of filtered P$_n$ methods in problems with heavy streaming. We used angular adaptivity in conjuction with FP$_n$ and this allowed us to easily build spatially dependent filter strengths that filter heavily near discontinuities in space/angle, while retaining their spectral order in smooth regions. This is in an attempt to produce fast, ray-effect free solutions to problems with heavy streaming and/or voids. 

We found that for problems without heavy streaming, (i.e., those with some smoothness), adaptive P$_n$ and adaptive FP$_n$ methods can perform better in NDOFs applied and runtime than uniform discretisations. In particular, care must be taken with FP$_n$ methods to not filter ``too much''. The introduction of a spatially dependent filter however made this process more robust, and our spatially dependent filter always outperformed a constant filter value. For problems with heavy streaming, we found our adaptive P$_n$ methods at least performed better than uniform P$_n$, or unsurprisingly were too poorly conditioned to solve. Surprisingly we found our adaptive FP$_n$ methods highly performant in streaming problems, with the spatially dependent filter giving performance comparable to adapted P$^0$ discretisations up to a reasonable order (approx. FP$_{29}$). In particular, we found close to fixed iteration counts with angular refinement. 

This gives a spherical harmonics method that is very close to ``scalable''. Of course the last barrier to true scalability in spherical harmonics methods is applying/computing inflow/outflow/BCs. At high-order the (at best) $\mathcal{O}(n^2)$ nature of these computations will always dominate, but our adaptive process at least allows the number/size of these computations to be minimised. We also found that using a spatially dependent filter in a goal-based FP$_n$ simulations with a pure vacuum allowed us to get non-zero responses in our functionals at the end of a duct with aspect ratio of 1:100, even with low-order FP$_n$ discretisations. This triggered refinement in our spherical harmonics in a problem where previous work \cite{Dargaville2019} has found significant difficulties in developing robust goal-based metrics.

One fundamental disadvantage of all FP$_n$ methods presented, both in this work and in the literature is picking a filter strength for a given problem. This work showed that allowing an adaptive process to make this filter strength spatially dependent makes the initial choice for a filter value less important, and helps reduce the risk of over-filtering. Indeed, based on this work we would always recommend running with a spatially dependent filter, as it improved the convergence in all the cases we investigated. We believe this work shows that even without $\mathcal{O}(n)$ scaling in angle size, adaptive filtered spherical harmonics are a powerful tool for fast, ray-effect free solutions for problems that can include pure streaming. 
%~~~~~~~~~~~~~~~~~~~~~~~~~~~~~~~~~~~~
\section*{Acknowledgements}
The authors would like to acknowledge the support of the EPSRC through the funding of the EPSRC grant EP/P013198/1.

%~~~~~~~~~~~~~~~~~~~~~~~~~~~~~~~~~~~~
%~~~~~~~~~~~~~~~~~~~~~~~~~~~~~~~~~~~~
%~~~~~~~~~~~~~~~~~~~~~~~~~~~~~~~~~~~~



%% References

\section*{References}
\bibliographystyle{model1-num-names}
\bibliography{bib_library}

\begin{thebibliography}{32}
\expandafter\ifx\csname natexlab\endcsname\relax\def\natexlab#1{#1}\fi
\providecommand{\bibinfo}[2]{#2}
\ifx\xfnm\relax \def\xfnm[#1]{\unskip,\space#1}\fi
%Type = Article
\bibitem[{Buchan et~al.(2008)Buchan, Pain, Eaton, Smedley-Stevenson, and
  Goddard}]{Buchan2008244}
\bibinfo{author}{A.~G. Buchan}, \bibinfo{author}{C.~C. Pain},
  \bibinfo{author}{M.~D. Eaton}, \bibinfo{author}{R.~P. Smedley-Stevenson},
  \bibinfo{author}{A.~J.~H. Goddard},
\newblock \bibinfo{title}{Self-adaptive spherical wavelets for angular
  discretisations of the boltzmann transport equation},
\newblock \bibinfo{journal}{Nucl.~Sci.~Eng.} \bibinfo{volume}{158}
  (\bibinfo{year}{2008}) \bibinfo{pages}{244--263}.
%Type = Article
\bibitem[{Goffin et~al.(2014)Goffin, Buchan, Belme, Pain, Eaton, Smith, and
  Smedley-Stevenson}]{Goffin2014}
\bibinfo{author}{M.~A. Goffin}, \bibinfo{author}{A.~G. Buchan},
  \bibinfo{author}{A.~C. Belme}, \bibinfo{author}{C.~C. Pain},
  \bibinfo{author}{M.~D. Eaton}, \bibinfo{author}{P.~N. Smith},
  \bibinfo{author}{R.~P. Smedley-Stevenson},
\newblock \bibinfo{title}{{Goal-based angular adaptivity applied to the
  spherical harmonics discretisation of the neutral particle transport
  equation}},
\newblock \bibinfo{journal}{Ann. Nucl. Energy} \bibinfo{volume}{71}
  (\bibinfo{year}{2014}) \bibinfo{pages}{60--80}.
%Type = Article
\bibitem[{Goffin et~al.(2015)Goffin, Buchan, Dargaville, Pain, Smith, and
  Smedley-Stevenson}]{Goffin2015}
\bibinfo{author}{M.~A. Goffin}, \bibinfo{author}{A.~G. Buchan},
  \bibinfo{author}{S.~Dargaville}, \bibinfo{author}{C.~C. Pain},
  \bibinfo{author}{P.~N. Smith}, \bibinfo{author}{R.~P. Smedley-Stevenson},
\newblock \bibinfo{title}{Goal-based angular adaptivity applied to a
  wavelet-based discretisation of the neutral particle transport equation},
\newblock \bibinfo{journal}{Journal of Computational Physics}
  \bibinfo{volume}{281} (\bibinfo{year}{2015}) \bibinfo{pages}{1032--1062}.
%Type = Phdthesis
\bibitem[{Goffin(2015)}]{Goffin2015a}
\bibinfo{author}{M.~Goffin}, \bibinfo{title}{Goal-based adaptive methods
  applied to the spatial and angular dimensions of the transport equation},
  Ph.D. thesis, Imperial College London, \bibinfo{year}{2015}.
%Type = Article
\bibitem[{Adam et~al.(2016)Adam, Buchan, Piggott, Pain, Hill, and
  Goffin}]{Adam2016a}
\bibinfo{author}{A.~Adam}, \bibinfo{author}{A.~G. Buchan},
  \bibinfo{author}{M.~D. Piggott}, \bibinfo{author}{C.~C. Pain},
  \bibinfo{author}{J.~Hill}, \bibinfo{author}{M.~A. Goffin},
\newblock \bibinfo{title}{Adaptive {Haar} wavelets for the angular
  discretisation of spectral wave models},
\newblock \bibinfo{journal}{Journal of Computational Physics}
  \bibinfo{volume}{305} (\bibinfo{year}{2016}) \bibinfo{pages}{521--538}.
%Type = Phdthesis
\bibitem[{Adam(2016)}]{Adam2016}
\bibinfo{author}{A.~Adam}, \bibinfo{title}{Finite element, adaptive spectral
  wave modelling}, Ph.D. thesis, Imperial College London, \bibinfo{year}{2016}.
%Type = Article
\bibitem[{Soucasse et~al.(2017)Soucasse, Dargaville, Buchan, and
  Pain}]{Soucasse2017}
\bibinfo{author}{L.~Soucasse}, \bibinfo{author}{S.~Dargaville},
  \bibinfo{author}{A.~G. Buchan}, \bibinfo{author}{C.~C. Pain},
\newblock \bibinfo{title}{A goal-based angular adaptivity method for thermal
  radiation modelling in non grey media},
\newblock \bibinfo{journal}{Journal of Quantitative Spectroscopy and Radiative
  Transfer} \bibinfo{volume}{200} (\bibinfo{year}{2017})
  \bibinfo{pages}{215--224}.
%Type = Article
\bibitem[{Dargaville et~al.(2019)Dargaville, Buchan, Smedley-Stevenson, Smith,
  and Pain}]{Dargaville2019}
\bibinfo{author}{S.~Dargaville}, \bibinfo{author}{A.~G. Buchan},
  \bibinfo{author}{R.~P. Smedley-Stevenson}, \bibinfo{author}{P.~N. Smith},
  \bibinfo{author}{C.~C. Pain},
\newblock \bibinfo{title}{Scalable angular adaptivity for {Boltzmann}
  transport},
\newblock \bibinfo{journal}{arXiv:1901.04929 [physics]}
  (\bibinfo{year}{2019}). \bibinfo{note}{ArXiv: 1901.04929}.
%Type = Article
\bibitem[{McClarren and Hauck(2010)}]{McClarren2010}
\bibinfo{author}{R.~G. McClarren}, \bibinfo{author}{C.~D. Hauck},
\newblock \bibinfo{title}{Robust and accurate filtered spherical harmonics
  expansions for radiative transfer},
\newblock \bibinfo{journal}{Journal of Computational Physics}
  \bibinfo{volume}{229} (\bibinfo{year}{2010}) \bibinfo{pages}{5597--5614}.
%Type = Article
\bibitem[{Radice et~al.(2013)Radice, Abdikamalov, Rezzolla, and
  Ott}]{Radice2013}
\bibinfo{author}{D.~Radice}, \bibinfo{author}{E.~Abdikamalov},
  \bibinfo{author}{L.~Rezzolla}, \bibinfo{author}{C.~D. Ott},
\newblock \bibinfo{title}{A new spherical harmonics scheme for
  multi-dimensional radiation transport {I}. {Static} matter configurations},
\newblock \bibinfo{journal}{Journal of Computational Physics}
  \bibinfo{volume}{242} (\bibinfo{year}{2013}) \bibinfo{pages}{648--669}.
%Type = Article
\bibitem[{Frank et~al.(2016)Frank, Hauck, and Kuepper}]{Frank2016}
\bibinfo{author}{M.~Frank}, \bibinfo{author}{C.~Hauck},
  \bibinfo{author}{K.~Kuepper},
\newblock \bibinfo{title}{Convergence of filtered spherical harmonic equations
  for radiation transport},
\newblock \bibinfo{journal}{Commun. Math. Sci} \bibinfo{volume}{14}
  (\bibinfo{year}{2016}) \bibinfo{pages}{1443--1465}.
%Type = Phdthesis
\bibitem[{Laboure(2016)}]{Laboure2016b}
\bibinfo{author}{V.~M. Laboure}, \bibinfo{title}{Improved {Fully}-{Implicit}
  {Spherical} {Harmonics} {Methods} for {First} and {Second} {Order} {Forms} of
  the {Transport} {Equation} {Using} {Galerkin} {Finite} {Element}},
  \bibinfo{type}{Thesis}, \bibinfo{year}{2016}.
%Type = Article
\bibitem[{Laboure et~al.(2016)Laboure, McClarren, and Hauck}]{Laboure2016}
\bibinfo{author}{V.~M. Laboure}, \bibinfo{author}{R.~G. McClarren},
  \bibinfo{author}{C.~D. Hauck},
\newblock \bibinfo{title}{Implicit {Filtered} {PN} for {High}-{Energy}
  {Density} {Thermal} {Radiation} {Transport} using {Discontinuous} {Galerkin}
  {Finite} {Elements}},
\newblock \bibinfo{journal}{Journal of Computational Physics}
  \bibinfo{volume}{321} (\bibinfo{year}{2016}) \bibinfo{pages}{624--643}.
  \bibinfo{note}{ArXiv: 1601.08242}.
%Type = Article
\bibitem[{Laiu and Hauck(2018)}]{Laiu2018}
\bibinfo{author}{M.~P. Laiu}, \bibinfo{author}{C.~D. Hauck},
\newblock \bibinfo{title}{Positivity {Limiters} for {Filtered} {Spectral}
  {Approximations} of {Linear} {Kinetic} {Transport} {Equations}},
\newblock \bibinfo{journal}{Journal of Scientific Computing}
  (\bibinfo{year}{2018}).
%Type = Article
\bibitem[{Hughes et~al.(1998)Hughes, Feijóo, Mazzei, and
  Quincy}]{hughes_variational_1998}
\bibinfo{author}{T.~J.~R. Hughes}, \bibinfo{author}{G.~R. Feijóo},
  \bibinfo{author}{L.~Mazzei}, \bibinfo{author}{J.-B. Quincy},
\newblock \bibinfo{title}{The variational multiscale method—a paradigm for
  computational mechanics},
\newblock \bibinfo{journal}{Computer Methods in Applied Mechanics and
  Engineering} \bibinfo{volume}{166} (\bibinfo{year}{1998})
  \bibinfo{pages}{3--24}.
%Type = Article
\bibitem[{Hughes et~al.(2006)Hughes, Scovazzi, Bochev, and
  Buffa}]{hughes_multiscale_2006}
\bibinfo{author}{T.~J.~R. Hughes}, \bibinfo{author}{G.~Scovazzi},
  \bibinfo{author}{P.~B. Bochev}, \bibinfo{author}{A.~Buffa},
\newblock \bibinfo{title}{A multiscale discontinuous galerkin method with the
  computational structure of a continuous galerkin method},
\newblock \bibinfo{journal}{Computer Methods in Applied Mechanics and
  Engineering} \bibinfo{volume}{195} (\bibinfo{year}{2006})
  \bibinfo{pages}{2761--2787}.
%Type = Phdthesis
\bibitem[{Candy(2008)}]{candy_subgrid_2008}
\bibinfo{author}{A.~S. Candy}, \bibinfo{title}{Subgrid scale modelling of
  transport processes.}, \bibinfo{type}{Thesis or dissertation}, Imperial
  College London, \bibinfo{year}{2008}.
%Type = Article
\bibitem[{Buchan et~al.(2010)Buchan, Candy, Merton, Pain, Hadi, Eaton, Goddard,
  Smedley-Stevenson, and Pearce}]{buchan_inner-element_2010}
\bibinfo{author}{A.~G. Buchan}, \bibinfo{author}{A.~S. Candy},
  \bibinfo{author}{S.~R. Merton}, \bibinfo{author}{C.~C. Pain},
  \bibinfo{author}{J.~I. Hadi}, \bibinfo{author}{M.~D. Eaton},
  \bibinfo{author}{A.~J.~H. Goddard}, \bibinfo{author}{R.~P.
  Smedley-Stevenson}, \bibinfo{author}{G.~J. Pearce},
\newblock \bibinfo{title}{The inner-element subgrid scale finite element method
  for the boltzmann transport equation},
\newblock \bibinfo{journal}{Nuclear science and engineering}
  \bibinfo{volume}{164} (\bibinfo{year}{2010}) \bibinfo{pages}{105--121}.
%Type = Article
\bibitem[{Ackroyd and Wilson(1986)}]{Ackroyd1986}
\bibinfo{author}{R.~T. Ackroyd}, \bibinfo{author}{W.~E. Wilson},
\newblock \bibinfo{title}{Discontinuous finite elements for neutron transport
  analysis},
\newblock \bibinfo{journal}{Progress in Nuclear Energy} \bibinfo{volume}{18}
  (\bibinfo{year}{1986}) \bibinfo{pages}{39--44}.
%Type = Article
\bibitem[{Ackroyd and Wilson(1988)}]{Ackroyd1988}
\bibinfo{author}{R.~T. Ackroyd}, \bibinfo{author}{W.~E. Wilson},
\newblock \bibinfo{title}{{Composite finite element solutions for neutron
  transport}},
\newblock \bibinfo{journal}{{Ann.~Nucl.~Energy}} \bibinfo{volume}{15}
  (\bibinfo{year}{1988}) \bibinfo{pages}{397--419}.
%Type = Phdthesis
\bibitem[{Park(2006)}]{Park2006}
\bibinfo{author}{H.~Park}, \bibinfo{title}{Coupled {Space}-{Angle} {Adaptivity}
  and {Goal}-{Oriented} {Error} {Control} for {Radiation} {Transport}
  {Calculations}}, Ph.D. thesis, Georgia Institute of Technology,
  \bibinfo{year}{2006}.
%Type = Article
\bibitem[{Park and de~Oliveira(2009)}]{Park2009a}
\bibinfo{author}{H.~Park}, \bibinfo{author}{C.~R.~E. de~Oliveira},
\newblock \bibinfo{title}{Coupled {Space}-{Angle} {Adaptivity} for {Radiation}
  {Transport} {Calculations}},
\newblock \bibinfo{journal}{Nuclear Science and Engineering}
  \bibinfo{volume}{161} (\bibinfo{year}{2009}) \bibinfo{pages}{216--234}.
%Type = Inproceedings
\bibitem[{Rupp et~al.(2011)Rupp, Grasser, and Jungel}]{Rupp2011a}
\bibinfo{author}{K.~Rupp}, \bibinfo{author}{T.~Grasser},
  \bibinfo{author}{A.~Jungel},
\newblock \bibinfo{title}{Adaptive variable-order spherical harmonics expansion
  of the {Boltzmann} {Transport} {Equation}},
\newblock in: \bibinfo{booktitle}{2011 {International} {Conference} on
  {Simulation} of {Semiconductor} {Processes} and {Devices} ({SISPAD})}, pp.
  \bibinfo{pages}{151--154}.
%Type = Article
\bibitem[{Safarzadeh et~al.(2015)Safarzadeh, Shirani, and
  Minuchehr}]{Safarzadeh2015}
\bibinfo{author}{O.~Safarzadeh}, \bibinfo{author}{A.~S. Shirani},
  \bibinfo{author}{A.~Minuchehr},
\newblock \bibinfo{title}{Hybrid space–angle adaptivity for whole-core
  particle transport calculations},
\newblock \bibinfo{journal}{Annals of Nuclear Energy} \bibinfo{volume}{80}
  (\bibinfo{year}{2015}) \bibinfo{pages}{254--260}.
%Type = Article
\bibitem[{Buchan et~al.(2011)Buchan, Merton, Pain, and
  Smedley-Stevenson}]{Buchan2011}
\bibinfo{author}{A.~G. Buchan}, \bibinfo{author}{S.~R. Merton},
  \bibinfo{author}{C.~C. Pain}, \bibinfo{author}{R.~P. Smedley-Stevenson},
\newblock \bibinfo{title}{Riemann boundary conditions for the {Boltzmann}
  transport equation using arbitrary angular approximations},
\newblock \bibinfo{journal}{Annals of Nuclear Energy} \bibinfo{volume}{38}
  (\bibinfo{year}{2011}) \bibinfo{pages}{1186--1195}.
%Type = Article
\bibitem[{Mohlenkamp(1999)}]{Mohlenkamp1999}
\bibinfo{author}{M.~J. Mohlenkamp},
\newblock \bibinfo{title}{A fast transform for spherical harmonics},
\newblock \bibinfo{journal}{Journal of Fourier Analysis and Applications}
  \bibinfo{volume}{5} (\bibinfo{year}{1999}) \bibinfo{pages}{159--184}.
%Type = Article
\bibitem[{Lessig et~al.(2012)Lessig, de~Witt, and Fiume}]{Lessig2012}
\bibinfo{author}{C.~Lessig}, \bibinfo{author}{T.~de~Witt},
  \bibinfo{author}{E.~Fiume},
\newblock \bibinfo{title}{Efficient and accurate rotation of finite spherical
  harmonics expansions},
\newblock \bibinfo{journal}{Journal of Computational Physics}
  \bibinfo{volume}{231} (\bibinfo{year}{2012}) \bibinfo{pages}{243--250}.
%Type = Article
\bibitem[{Dargaville et~al.(2015)Dargaville, Goffin, Buchan, Pain,
  Smedley-Stevenson, Smith, and Gorman}]{Dargaville_2014}
\bibinfo{author}{S.~Dargaville}, \bibinfo{author}{M.~A. Goffin},
  \bibinfo{author}{A.~G. Buchan}, \bibinfo{author}{C.~C. Pain},
  \bibinfo{author}{R.~P. Smedley-Stevenson}, \bibinfo{author}{P.~N. Smith},
  \bibinfo{author}{G.~Gorman},
\newblock \bibinfo{title}{Solving the boltzmann transport equation with
  multigrid and adaptive space/angle discretisations},
\newblock \bibinfo{journal}{Annals of Nuclear Energy} \bibinfo{volume}{86}
  (\bibinfo{year}{2015}) \bibinfo{pages}{99--107}.
%Type = Article
\bibitem[{Adigun et~al.(2018)Adigun, Buchan, Adam, Dargaville, Goffin, and
  Pain}]{Adigun2018}
\bibinfo{author}{B.~J. Adigun}, \bibinfo{author}{A.~G. Buchan},
  \bibinfo{author}{A.~Adam}, \bibinfo{author}{S.~Dargaville},
  \bibinfo{author}{M.~A. Goffin}, \bibinfo{author}{C.~C. Pain},
\newblock \bibinfo{title}{A {Haar} wavelet method for angularly discretising
  the {Boltzmann} transport equation},
\newblock \bibinfo{journal}{Progress in Nuclear Energy} \bibinfo{volume}{108}
  (\bibinfo{year}{2018}) \bibinfo{pages}{295--309}.
%Type = Article
\bibitem[{Buchan and Pain(2016)}]{Buchan2016}
\bibinfo{author}{A.~G. Buchan}, \bibinfo{author}{C.~C. Pain},
\newblock \bibinfo{title}{An efficient space-angle subgrid scale discretisation
  of the neutron transport equation},
\newblock \bibinfo{journal}{Annals of Nuclear Energy} \bibinfo{volume}{94}
  (\bibinfo{year}{2016}) \bibinfo{pages}{440--450}.
%Type = Article
\bibitem[{Ragusa et~al.(2012)Ragusa, Guermond, and Kanschat}]{Ragusa2012}
\bibinfo{author}{J.~C. Ragusa}, \bibinfo{author}{J.~L. Guermond},
  \bibinfo{author}{G.~Kanschat},
\newblock \bibinfo{title}{A robust {SN}-{DG}-approximation for radiation
  transport in optically thick and diffusive regimes},
\newblock \bibinfo{journal}{Journal of Computational Physics}
  \bibinfo{volume}{231} (\bibinfo{year}{2012}) \bibinfo{pages}{1947--1962}.
%Type = Article
\bibitem[{Brunner(2002)}]{Brunner2002}
\bibinfo{author}{T.~A. Brunner},
\newblock \bibinfo{title}{Forms of approximate radiation transport},
\newblock \bibinfo{journal}{Sandia report}  (\bibinfo{year}{2002}).

\end{thebibliography}






% \end{thebibliography}

\end{document}